%% file: paper-popsyn-xxx.tex
\documentclass[amstex,apj]{emulateapj}
\usepackage{graphicx}
\usepackage{amsmath}
\usepackage{hyperref}           
\newcommand\unit[1]{\,{\rm #1}}
\newcommand\mynmin{20}
\newcommand\myratio{5\times 10^6}
\newcommand\abbrvChrisSpinup{BTRS}

\newcommand\draftonly[1]{}
\newcommand\editremark[1]{{}}

\newcommand\embrace[1]{{[#1]}}
\begin{document}

\newcommand{\nsns}{NS$-$NS\ }

\title{Short Gamma-Ray Bursts and Binary Mergers 
  in Spiral and Elliptical Galaxies: 
   Redshift Distribution and Hosts}  
\author{R.\ O'Shaughnessy}
\affil{Northwestern University, Department of Physics and Astronomy,
  2145 Sheridan Road, Evanston, IL 60208, USA}
\email{oshaughn@northwestern.edu}
\author{ K.\ Belczynski}
\affil{Tombaugh Fellow, New Mexico State University, Las
   Cruces, New Mexico, 88003, USA}
\author{V.\ Kalogera}
\affil{Northwestern University, Department of Physics and Astronomy,
  2145 Sheridan Road, Evanston, IL 60208, USA}

\begin{abstract}
To critically assess the binary compact object merger model for short
gamma ray bursts (GRBs) -- specifically, to test whether the short GRB rates, redshift distribution and 
host galaxies are consistent with current theoretical predictions -- it is necessary to 
examine models that account for the high-redshift, heterogeneous universe (accounting for both spirals and ellipticals). We present an investigation of predictions produced from a very large database of first-principle population synthesis calculations for 
binary compact mergers with neutron stars (NS) and black holes (BH), that sample a seven-dimensional space for binaries and their evolution. 
We further link these predictions to
 (i) the star formation history of the universe, (ii) a
heterogeneous population of star-forming galaxies, including spirals and ellipticals, 
and (iii) a simple selection model for bursts based on flux-limited detection. 
We impose a number of constraints on the model predictions at different quantitative levels: short GRB rates and redshift measurements, and, for NS-NS, the current empirical estimates 
of Galactic merger rates derived from the observed sample of close binary pulsars. 
Because of the relative weakness of these observational
constraints (due to small samples and measurement uncertainties), 
we find a small, but still substantial, fraction  of models are
agreement with available short GRB and binary pulsar observations,
both when we assume short GRB mergers are associated with NS-NS mergers and when we assume they are associated with BH-NS mergers.  
Notably, we do not need to introduce artificial models with {\em exclusively} long delay times.
Most commonly models produce mergers preferentially in spiral
galaxies, in fact predominantly so, if short GRBs arise from
NS-NS mergers alone. On the other hand, typically BH-NS mergers 
can also occur in elliptical galaxies (for some models, even preferentially), in agreement
with existing observations.  As one would expect,
model universes where present-day BH-NS binary mergers occur preferentially in
elliptical galaxies necessarily include a significant fraction of binaries with long delay times between birth and merger (often $O(10{\rm
  Gyr})$); unlike previous attempts to fit observations, these long
delay times arise naturally as properties of our model populations.
Though long delays occur, almost all of our models (both \emph{a
  priori} and constrained) predict that a higher proportion of short
GRBs should occur at moderate to high redshift (e.g., $z>1$) than has
presently been observed, in agreement with recent observations which
suggest a strong selection bias towards successful follow-up of
low-redshift short GRBs.
Finally, if we adopt plausible priors on the fraction of BH-NS mergers
with appropriate combination of spins and masses to produce
a short GRB event based on 
\citet{ChrisSpinup2007}, then at best only a small fraction of BH-NS models
could be consistent with all {\em current} available data, whereas NS-NS models do so more naturally.
\end{abstract}
\keywords{Stars: Binaries: Close; Gamma-ray bursts}

\maketitle

\section{Introduction}
The afterglows of several  short-hard gamma ray bursts (GRBs) have
recently been localized on the sky, allowing
reasonably precise determination 
of their hosts, redshifts, and energetics \citep[see,
e.g.,][]{berger-manyfainthosts2006,Berger2006talk,grb-050709-discovery,2005Natur.437..845F}.
The  energetics, presence in both old and star forming  
host galaxies, absence of supernovae afterglow characteristics, and in some cases 
apparent host offsets  of these bursts  seem qualitatively consistent
with the \emph{merger hypothesis} (MH): the 
notion that most short-hard bursts arise from the disruption of a neutron star (NS) in either a NS-NS or black
hole-NS binary  (see, e.g, \citet{2005ApJ...630L.165L} and references therein;
this GRB model was first presented by \cite{1986ApJ...308L..43P}).
While other populations, such as bursts from magnetars, may
contribute to the total short GRB event rate, 
the fraction of  GRBs produced by nearby magnetars is believed to be small
\citep[see,
e.g.,][]{2006ApJ...640..849N,2006MNRAS.365..885P,2005MNRAS.362L...8L}.
Furthermore, empirical and theoretical estimates for compact object
merger rates based on studies of the Milky Way \citep[see,
e.g.,][]{Chunglee-nsns-1,PSconstraints,PSmoreconstraints,StarTrack,2006astro.ph.12032B,2004ApJ...614L.137K,NagamineTwoComponentSFR2006} are roughly consistent with 
BATSE and Swift observations 
\citep{GuettaPiran2006,GuettaPiran2005,Ando2004}.


The number of observed radio pulsars with neutron star
companions  can provide a
robust  quantitative test of the MH.  For example, using
well-understood selection  effects and fairly minimal population
modeling (i.e., a luminosity 
function and a beaming correction factor), \citet{Chunglee-nsns-1}
developed a statistical method to determine which double neutron star coalescence
rates were  consistent with NS-NS seen in the
Milky Way.
However, in distinct contrast to NS-NS in the Milky Way,  little
is known directly about the short GRB  spatial or luminosity
distribution.

Short GRBs are still detected quite infrequently (i.e, a handful of
detections  per
year for Swift); sufficient statistics are not available for a 
robust nonparametric estimate of their 
distribution in redshift $z$ and peak luminosity $L$.   To make good use of the observed
$(z,L)$ data, we must fold in fairly
strong prior assumptions about the  two-dimensional density $d^3N/dtdLdz
(L,z)$.  
Typically, these priors are constructed by convolving the star
formation history of the universe with a hypothesized distribution for
the ``delay time'' between the short GRB progenitor's birth and
the GRB event, as well as with an effective (detection- and
angle-averaged) luminosity distribution for short GRBs.   
Observations are thus interpreted as constraints on the space of
models, rather than as direct measurements of the $z,L$ distribution
\citep{Ando2004,GuettaPiran2005,GuettaPiran2006,2005astro.ph..9891G}.  
A similar technique has been applied with considerable success to long
GRB observations
\citep[see,e.g.,][and references therein]{PorcianiMadau,GuettaPiran2005,1999ApJ...523L.117S,1999ApJ...516..559C}: as expected from a supernovae origin,
the long GRB rate is consistent with the star formation history of the
universe.
%
And within the context of \emph{specific assumptions} about the
merger delay time distribution and star formation history of the
universe (i.e., $dn/dt\propto 1/t$ and homogeneous through all space,
respectively),   \cite{2005astro.ph..9891G} and 
\cite{Nakar} have compared whether their set of models can produce results
statistically consistent with observations.  Among other things they
conclude that, 
within these conventional assumptions, the merger  model seems
inconsistent with the data.

These previous predictions assume homogeneous star forming
conditions throughout the universe, with rate proportional to the
observed time-dependent star formation rate 
(as given by, for example, \citet{NagamineTwoComponentSFR2006} and references therein).
In reality, however, the universe is markedly heterogeneous as well as
time-dependent; for example, large elliptical galaxies form most of
their stars early on.  
%
Similarly, predictions for the delay time distribution and indeed the total
number of compact binaries depend strongly on the
assumptions entering into population synthesis simulations.   These
simulations evolve a set of representative stellar systems using the
best parameterized recipies for weakly-constrained (i.e., supernovae) or computationally
taxing (i.e., stellar evolution) features of single and binary stellar
evolution.   By changing the specific assumptions used in these
recipies, physical predictions such as the NS-NS merger rate can vary
by a few orders of magnitude \citep[see,e.g.][and references
therein]{KNST}. In particular, certain model
parameters may allow much better agreement with observations.

In this study we examine predictions based on a large database of 
conventional binary population synthesis models: two sets of 500
concrete pairs of simulations 
(\S\ref{sec:results:full}), 
each pair of simulations modeling elliptical and spiral
galaxies respectively.\footnote{%
Because simulations that produce many BH-NS mergers need not
  produce many NS-NS mergers and vice-versa, we perform two
  independent sets of 500 pairs of simulations, each set designed to
  explore the properties of one particular merger type (i.e, BH-NS or
  NS-NS).  The statistical biases motivating this substantial increase
in computational effort are discussed in the Appendix.
}  In order to make predictions regarding the elliptical-to-spiral rate ratio for 
binary mergers, we adopt a two-component model for the star
formation history of the universe (\S\ref{sec:sfr}).
Our predictions include many models which agree with all existing (albeit weak)
observational constraints we could reliably impose.  Specifically,
many models (roughly half of all examined) reproduce the observed short-GRB
redshift distribution,  when 
we assume either NS-NS or  BH-NS progenitors.  Fewer  NS-NS
 models (roughly a tenth of all examined) can
reproduce both the short GRB redshift distribution and the NS-NS
 merger rate in spiral-type galaxies, as inferred from
observations of double pulsars seen in the Milky Way \citep[see,e.g.][]{Chunglee-nsns-1}.
We extensively describe the properties of those simulations
which reproduce observations (\S\ref{sec:results:full}): the redshift
distribution, the fraction of bursts with spiral hosts, and the
expected detection 
rate (given a fixed minimum burst luminosity). We present our
conclusions in section \ref{sec:conclude}.


\section{Gamma ray bursts:  Searches and Observations}
\label{sec:grb}
\subsection{Emission and detection models}
To compare the predictions of population synthesis
calculations with  the observed sample of short GRBs, we must estimate
the probability of detecting a given burst.   We therefore introduce 
 (i) a GRB emission model consisting of an effective
luminosity function for the isotropic energy emitted, to determine the
relative probability of various peak fluxes,  and a spectral
model, for K-corrections to observed peak fluxes, and (ii) a 
detection model introducing a fixed peak-flux detection threshold.    
%
Overall we limit attention to relatively simple models for both GRB emission and
detection. 
Specifically, we assume telescopes such as BATSE and Swift detect all
sources in their time-averaged field of view  ($\approx 2\pi$ and $1.4$ steradians,
respectively;  corresponding to a detector-orientation correction factor
  $f_d$ given by  $1/f_d=1/2$ and $1.4/4\pi$) 
with peak fluxes at the detector $F_d$
greater than some fixed threshold 
of  
$F_d= 1 {\rm ph}\, {\rm sec}^{-1}
{\rm cm}^{-2}$ in $50$ to $300$ keV 
\citep[see,e.g.][]{1997.astro-ph..9712091}. 
We note that Swift's triggering mechanism is more complex (Gehrels, private
communication) and appears biased against detections of short bursts;
for this reason, BATSE results and detection strategies will be emphasized heavily in what follows.

Similarly, though  observations of  short gamma ray bursts
reveal a variety of  spectra
\citep[see,e.g.][keeping in mind the observed peak energy is
redshifted]{2004AA...422L..55G}, and though this variety can have
significant implications for the detection of \emph{moderate-redshift}
($z>1$) bursts, for the purposes of this paper
we assume all short gamma ray bursts possess a pure power-law spectrum 
$
F_\nu\propto \nu^{-\alpha}
$
with $\alpha=-0.5$.   
Though several authors such as \citet{Ando2004} and
\citet{2001ApJ...552...36S}  have employed more realistic bounded
spectra,  similar pure power-law spectra have been
applied to interpret low-redshift observations in previous theoretical data
analysis efforts: 
\citet{Nakar} uses precisely this spectral index;
\citet{GuettaPiran2006} use $\alpha=-0.1$.\footnote{In reality,
  however, a break in the spectrum is often observed, redshifted into
  the detection band.  Under these circumstances, the K-correction can
  play a significant role in detectability.
}

Because our spectral model is manifestly unphysical outside our
detection band ($50-300$ keV), we cannot  relate observed, redshifted fluxes to
total luminosity.  Instead, we characterize the source's intrinsic
photon luminosity by the rate $\dot{N}=dN/dt_e$ at which it appears to produce $B=50-300$
keV photons isotropically in its rest frame, which we estimate from observed
 fluxes $F$ in this band via a K-correction: 
\begin{eqnarray}
\dot{N}&\equiv& F (4 \pi r^2)(1+z) k(z) \\
k(z) &\equiv&\frac{\int_{B} F_{\nu}d\nu/\nu}{\int_{B(1+z)} F_{\nu}
  d\nu/\nu}
 = (1+z)^{-0.5}
\end{eqnarray}
where $r(z)$ is the comoving distance at redshift $z$.
To give a sense of scale, a luminosity 
$L/(10^{47} {\rm  erg^{-1} s^{-1}})$ corresponds to a photon luminosity 
$\dot{N}/(4\times 10^{53} s^{-1}) $;  similarly, the characteristic distance
to which a photon flux can be seen is  $r_c\equiv \sqrt{N/4\pi F_d}\simeq
57 {\rm Mpc}(\dot{N}/4\times 10^{53} s^{-1})^{1/2}(F_d/1 {\rm cm}^{-2} s^{-1})^{-1/2}$.

\begin{figure}
\includegraphics[width=\columnwidth]{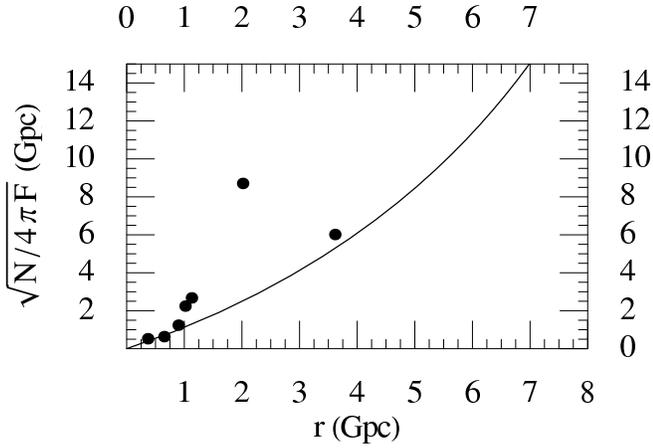}
\caption{\label{fig:grblf:criticalNdot}  Characteristic distance to a source
 $\sqrt{\dot{N}/4\pi F_d}$ versus its comoving distance.  
Points: 
Short bursts with well-defined redshifts (SH1; see Table \ref{tab:grbs}).
Solid line:
   The critical 
  characteristic distance $r_c(z)\equiv \sqrt{\dot{N}_d(z)/4\pi
    F_d}=r(z)\sqrt{(1+z)k(z)}$ versus comoving distance $r(z)$, for
  our simple power-law spectral model $F_\nu\propto \nu^{0.5}$.    Given 
  our assumptions, systems with fluxes $\dot{N}$
  corresponding to points above this curve can be seen at the
  earth with a band-limited detector in $50-300$ keV with peak flux
  $\ge F_d$.  
}
\end{figure}

Finally,  we assume that short GRBs possess an intrinsic
power-law peak flux distribution: that the peak fluxes seen by
detectors placed at a fixed distance but random orientation relative
to all short GRBs should either (i) be precisely zero, with
probability $1-1/f_b$ or (ii)  
collectively be power-law distributed, from some (unknown) minimum peak flux to
infinity, with some probability $1/f_b$.  [This defines 
$f_b$, the beaming correction factor, in terms of the relative
probabilities of a visible orientation.]  For convenience in calculation, we will convert this power-law
peak-flux distribution into its equivalent power-law photon rate
$\dot{N}$ distribution
\begin{equation}
P(>\dot{N}) \equiv \left\{
\begin{array}{ll}
 f_b^{-1}(\dot{N}/\dot{N}_{\rm  min})^{1-\beta} 
   & \text{if  } \dot{N}>\dot{N}_{\rm min} \\ 
 f_b^{-1} & \text{if } \dot{N}\le\dot{N}_{\rm min}
\end{array}
\right.
\end{equation}
where we assume $\beta=2$; this particular choice of the power-law
exponent is a good match to the observed BATSE peak-flux distribution
\citep[see,
e.g.][and references therein]{GuettaPiran2006,Nakar,Ando2004,2001ApJ...552...36S}.
The fraction of short bursts that are visible
at a redshift $z$ is thus $P(z)\equiv P(>\dot{N}_d)$, where
$\dot{N}_d$ is shorthand for $4 \pi r^2(1+z) k(z) F_d$.
Once again, these assumptions
correspond approximately to those previously published in the
literature; elementary extensions (for example, a wider
range of source luminosity distributions) 
 have been successfully applied to match the
observed BATSE flux distributions and Swift redshift-luminosity data
[e.g., in addition to the references mentioned previously, \citet{GuettaPiran2005}].

%



\subsection{GRB Observations}
\label{sec:grb:data}

While the above discussion summarizes the most critical selection
effects -- the conditions needed for GRB detection -- other more
subtle selection effects can significantly influence data interpretation.
Even assigning a burst to the ``short'' class uses a fairly
coarse  phenomenological classification [compare, e.g., the modern
spectral evolution classification of \cite{2006ApJ...643..266N}, the
machine-learning methods of
\cite{2003ApJ...582..320H}, and the original classification paper \cite{grbs:general:discovery:multiple-classes}];
alternate classifications produce  slightly but
significantly different populations \citep[see,e.g.][for a concrete,
much broader classification scheme]{DonLamb-modifiedClassifications2006}.
Additionally,  short GRB redshift measurements can be produced only after a \emph{second}
optical search, with its own strong  selection biases toward
low-redshift hosts
\citep[see,e.g.,][]{berger-manyfainthosts2006}.   

\input{publishable-table.tex}

To avoid controversy, we therefore assemble our list of short GRBs
from four previously-published compilations:
(i) \cite{berger-manyfainthosts2006}  (Table 1), which provides a
state-of-the-art Swift-dominated sample with relatively homogeneous
selection effects; (ii) \cite{DonLamb-modifiedClassifications2006}
(Table 8), a broader sample defined using an alternative short-long
classification;
 and finally (iii) \cite{berger-shortgrb-parameter-correlation-2007} and
 (iv) \cite{Gehrels-shortgrb-SwiftReview-Mid2007}
which cover the gaps between the first two and the present.  [We limit attention to bursts
seen  since 2005, so selection effects are fairly constant through the
observation interval.
  For similar reasons, we do not include the post-facto IPN galaxy
associations shown in  \cite{Nakar} (Table 1).]  This compilation
omits GRB 050911 discussed in
\cite{2006ApJ...637L..13P} but otherwise includes most proposed short GRB candidates.
As shown in Table \ref{tab:grbs}, the sample consists of 21
bursts; though  most   (15) have some redshift information,
only 11  have relatively well-determined redshifts.  
However, even among these 12 sources, some disagreement
exists regarding the correct host associations and redshifts of GRBs
060505 and  060502B \citep[see,e.g.,][]{berger-manyfainthosts2006}.

To make sure the many hidden uncertainties and selection biases are
explicitly yet compactly noted in subsequent calculations,  we
introduce a simple hierarchical classification 
for bursts seen since 2005: S$n$ represent the  
bursts detected only with Swift; SH$n$ the 
bursts seen either by Swift or HETE-II; $n=1$ corresponds to bursts
with well-determined redshifts; $n=2$ corrresponds to bursts with some
strong redshift constraints; and $n=3$ includes all bursts.

Starting in  May 2005, Swift detected 9
  short GRBs in a calendar year.  For the purposes of comparison, we
  will assume the Swift short GRB detection rate to be $R_{D,\rm Swift}=10 {\rm
  yr}^{-1}$; 
compensating for its partial sky coverage, this rate
corresponds to an all-sky event rate at earth of $f_{d,\rm Swift} R_{D,\rm
  Swift}\simeq 90 \unit{yr}^{-1}$.
However, in order to better account for the strong selection biases
apparently introduced by the Swift triggering mechanism against short
GRBs (Gehrels, private communication), we assume the rate of GRB events above this threshold at earth
to be much better represented by the BATSE  detection rate $R_{d,\rm
  BATSE}$ when corrected for detector sky coverage, namely
$f_{d,\rm BATSE}R_{D,\rm BATSE} = 170\unit{yr}^{-1}$
\citep{1999ApJS..122..465P}\footnote{ Section 2  of \citet{GuettaPiran2005}
  describes how this rate can be extracted from the BATSE catalog
  paper, taking into account time-dependent changes in the
  instrument's selection effects.
}.  
For similar reasons, in this paper we express detection and
sensitivity limits in a BATSE band (50-300 keV) rather than the Swift BAT band.



\subsection{Cumulative redshift distribution}
\begin{figure}
\includegraphics[width=\columnwidth]{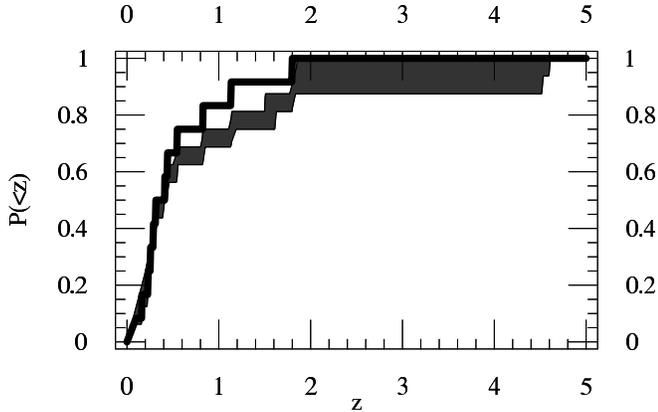}
\caption{\label{fig:data:grbs}  Cumulative redshift distribution of
  detected short GRBs.  The thick solid curve provides the cumulative
  distribution of well-constrained GRBs (i.e., the class SH1).  The
  shaded region indicates the range of cumulative distributions
  produced by assigning redshifts to the weakly-constrained  (i.e., the class SH2)
  in a manner consistent with the 
  constraints.  When only an upper or lower limit is available, we
  pick redshifts  using a uniform prior  for redshifts between 0 and 5.
}
\end{figure}

As \cite{Nakar} demonstrated and as described in detail
in \S\ref{sec:results:full}, the cumulative redshift distribution depends
very weakly on most parameters in the short GRB emission and detection
model (i.e., $f_b$, $f_d$, $\dot{N}$,  and $F_d$).  When sufficiently
many unbiased redshift measurements are available to estimate it, the
observed redshift distribution can stringently constrain models which
purport to reproduce it.
At present, however, only 11 reliable redshifts  are available,
leading to the cumulative redshift distribution shown in  Figure
\ref{fig:data:grbs} (thick solid line).   We can increase
this sample marginally by including more weakly-constrained sources. 
In Figure \ref{fig:data:grbs} (shaded region) we show several distributions consistent
with SH2, choosing redshifts uniformly from the intersection of
the region satisfying any constraints and $0<z<5$ (an interval which
encompasses all proposed short GRB redshifts).   
Because this larger sample includes a disproportionate number of
higher-redshift possibilities, 
the resulting cumulative redshift distributions still agree at
very low redshifts.

The small sample size seriously limits our ability to accurately
measure the  cumulative
distribution: given the sample size, a Kolmogorov-Smirnov 95\%
confidence interval includes \emph{any} distribution which deviates 
by less than $0.375$ from the observed cumulative distribution. 
Rather than account for all possibilities allowed by
  observations,
we will
accept any  model with maximum distance less than $0.375$ from
the cumulative redshift distribution for the well-known bursts (i.e.,
from the solid curve in in Figure \ref{fig:data:grbs}).

 By performing 
deep optical searches to identify hosts for unconstrained bursts,
\cite{berger-manyfainthosts2006}  have demonstrated that recent afterglow
studies are biased towards low redshift -- nearby galaxies are much
easier to detect optically than high-redshift hosts -- and that a
substantial population of high-redshift short bursts should exist.  This
high-redshift population becomes more  apparent when a few high-redshift
afterglows seen with HETE-II before 2005 are included; 
see \citet{DonLamb-modifiedClassifications2006} for details.

\subsection{Comparison with prior work}
\noindent \emph{Short GRB interpretation}:
Several previous efforts have been made  to test quantitative MH-based predictions
for the  host, redshift, luminosity, and age distributions
[\cite{2006.astro-ph..0606377,GuettaPiran2006,
Nakar,2005astro.ph..9891G,ADM:Blo99,ChrisShortGRBs,ADM:Per2002}].
However, many authors found puzzling discrepancies; most
notably, as has been emphasized by \citet{2005astro.ph..9891G,Nakar}
and by \citet{GuettaPiran2006} (by comparing redshift-luminosity
distributions to models) and as has seemingly been experimentally
  corroborated with GRB 060502B
\cite{2006ApJ...638..354B},  typical observed short GRBs appear to
occur  $\approx 
{\rm (1- few)
  \times Gyr}$ after
their progenitors' birth.  By contrast, most authors expect population
synthesis predicts a delay time distribution $dp/dt\propto 1/t$ (e.g.,
Piran 1992),
which was interpreted to imply that short delay times dominate, that  DCO mergers occur very soon after
birth, and that mergers observed on our  light cone predominantly
arise from recent star formation.  
Additionally, on the basis of the \emph{observed} redshift-luminosity
distribution alone, \citet{GuettaPiran2006} and \citet{Nakar} conclude
short GRB rates to be at least comparable to observed present-day NS-NS
merger rate in the Milky Way.  They both also note that a
large population of low-luminosity bursts (i.e., $L<10^{49}\,$erg)
would remain undetected, a possibility which may have some
experimental support: post-facto correlations between short GRBs and nearby galaxies
suggests the minimum luminosity of gamma ray bursts ($L_{min}$) could
be orders of magnitude lower 
\citep{Nakar,2005Natur.438..991T}. Such a large population would lead to a discrepancy between the two types of inferred rates.  
In summary, straightforward studies of the observed SHB sample suggest (i)
delay times and (ii) to a lesser extent rate densities are 
at least marginally and
possibly significantly incongruent with
the observed present-day Milky Way sample of double NS binaries, and
by extension the merger hypothesis \citep[cf. Sections 3.2 and 4 of][]{Nakar}.
%
A more recent study by \cite{berger-manyfainthosts2006} suggests that
high-redshift hosts may be significantly more difficult to identify
optically.  Using the relatively weak constraints they obtain
regarding the redshifts of previously-unidentified bursts, they
reanalyze the data to find  delay time
distributions  consistent with $dP/dt\propto 1/t$, as qualitatively
expected from detailed evolutionary simulations.

In all cases, however, these comparisons were based on elementary, semianalytic 
population models, with no prior on the relative likelihood of
different models: a model with a Gyr characteristic delay between
birth and merger was a priori as likely as $dP/dt\propto 1/t$.  For
this reason, our
study uses a large array of concrete population synthesis simulations,
in order to estimate the relative likelihood of different delay time
distributions. 



\noindent \emph{Population synthesis}: Earlier population synthesis
studies have explored similar subject matter, even including heterogeneous galaxy populations 
\citep[see,
e.g.][]{ChrisShortGRBs,ADM:de2005,ADM:Per2002,ADM:Blo99,1999ApJ...526..152F,2002ApJ...571..394B}.
These studies largely explored a single preferred model, in order to
produce what these authors expect as the 
\emph{most likely} predictions, such as for the offsets
expected from merging supernovae-kicked binaries and the likely gas content of
the circumburst environment.  Though preliminary short GRB
observations appeared to contain an overabundance of short GRBs
\citep{Nakar},  recent 
observational analyses such as \cite{berger-manyfainthosts2006} suggest
high-redshift bursts are also present, in qualitative agreement with the
detailed population synthesis study by \citet{ChrisShortGRBs}.  
The present study quantitatively reinforces this conclusion through carefully
reanalyzing the implications of short GRB observations, and
particularly through properly accounting for the small short GRB
sample size.

The extensive parameter study included here, however, bears closest
relation to a similar slightly smaller study in
\cite{2002ApJ...571..394B}, based on 30 population synthesis models.
Though intended for all GRBs, the range of predictions remain highly
pertinent for the short GRB population.
In most respects this earlier study was much broader than the present work: it
examined a much wider range of potential central engines (e.g., white
dwarf-black hole mergers) and extracted a wider range of predictions
(e.g.,  offsets from the central host).  The present paper, however, not only explores
a much larger set of population synthesis models ($\simeq$500)  -- including an
entirely new degree of freedom, the relative proportion of short GRBs hosted in
elliptical and spiral galaxies -- but also compares predictions
specifically against short GRB observations.




\section{Other Relevant Observations}
\subsection{Multicomponent star formation history}
\label{sec:sfr}

\begin{figure}
\includegraphics[width=\columnwidth]{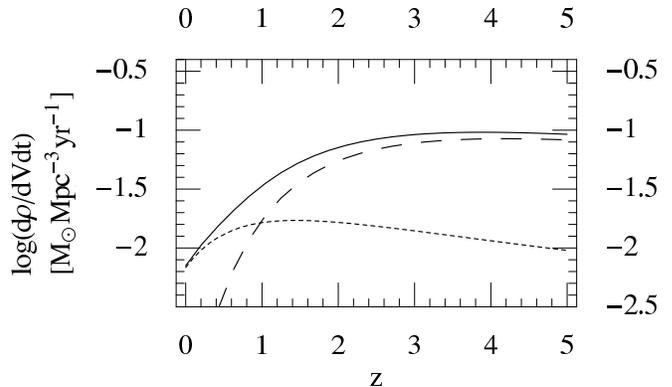}
\caption{\label{fig:sfr:observed} 
 Star formation history of the universe used in this paper.
Solid line:  Net star formation history implied by
  Eq. (\ref{eq:sfr:model:twocomponent}).
Dashed, dotted line: The star formation history due to  elliptical and
spiral galaxies, respectively.
}
\end{figure}

The star formation history of the universe has been extensively
explored through a variety of methods: extraglactic background light
modulo extinction 
\citep[see,e.g.,][and references
therein]{NagamineTwoComponentSFR2006,2004ApJ...615..209H}; direct galaxy
counts augmented by mass estimates \citep[see,e.g.][and references
therein]{2005ApJ...625..621B}; 
galaxy counts augmented with reconstructed star formation histories from their
 spectral energy distribution
 \citep[e.g.][]{Heavens,2006astro.ph..5060T,2006astro.ph..4554Y,2006astro.ph..4442H};
 and via more general composite methods \citep{2005JCAP...04..017S}.
Since all methods estimate the \emph{total} mass formed in stars from
some \emph{detectable} quantity, the result depends sensitively on the
assumed low-mass IMF and often on extinction.  However, as recently
demonstrated by \citet{sfr-Hopkins-review-2006} and \citet{NagamineTwoComponentSFR2006}, when observations are
interpreted in light of a common Chabrier IMF, observations  
largely converge upon a unique star-formation
rate  per unit comoving volume $\dot{\rho} = dM/dV dt$
bridging nearby and distant universe, as shown in Figure
\ref{fig:sfr:observed}.

Less clearly characterized in the literature are the \emph{components} of
the net star formation history $\dot{\rho}$: the history of star
formation in relatively well-defined subpopulations such as elliptical
and spiral galaxies.\footnote{Short GRBs have been associated with more refined
  morphological types, such as dwarf irregular galaxies.  For the
  purposes of this paper, these galaxies are sufficiently star forming
  to be ``spiral-like''. 
}  
For most of time, galaxies have existed in
relatively well-defined populations, with fairly little morphological
evolution outside of rare overdense regions \citep[see, e.g.][and
references therein]{2005ApJ...625..621B,2006astro.ph..4442H}.
Different populations possess dramatically different histories: the most
massive galaxies form most of their stars very early on
\citep[see,e.g.][]{2005ApJ...633L...9F} and hence at a characteristically lower
metallicity.  
%
Further, as has been extensively advocated by Kroupa \citep[see,
e.g.][and references
therein]{KroupaClusterAverageIMF2.7,2006astro.ph..4534F}
the most massive structures could conceivably form stars
through an entirely different collapse mechanism
(``starburst-mode'', driven for example by galaxy collisions and
capture) than the throttled mode relevant to disks of spiral galaxies
(``disk-mode''), resulting in particular in a different IMF.  


Both components significantly influence the present-day merger rate.  
For example, the initial mass function determines how many progenitors
of compact binaries are born from star-forming gas and thus are
available to evolve into merging BH-NS or NS-NS binaries.
Specifically, as shown in
detail in   \S
\ref{sec:popsyn} and particularly via Figure
\ref{fig:popsyn:distrib:epsilonlambda}, elliptical galaxies produce
roughly three times more high mass binaries per unit mass
than their spiral counterparts.
Additionally, as first recognized by
\citet{Regimbau2006-ellipticals}, even though elliptical galaxies are
quiescent now, the number of compact binaries formed in ellipticals decays
roughly \emph{logarithmically} with time (i.e., $dn/dt\propto 1/t$).
Therefore, due to the high 
star formation rate in elliptical-type galaxies $\sim 10\unit{Gyr}$
ago,    the star forming mass density $\delta \rho_e$ born in
ellipticals roughly $t_e\sim 10\unit{Gyr}$  ago  produces mergers at
a  rate density  $\sim \delta \rho_e /t_e$ that is often
comparable to or larger than  the rate density of mergers occurring
soon after their birth in spiral galaxies $\sim d\rho_s/dt$.

\subsubsection{Standard two-component model}
As a reference model we use the two-component star formation history
model presented by \citet{NagamineTwoComponentSFR2006}.  This model
consists of an early ``elliptical'' component and a fairly steady
``spiral'' component, with star formation rates given by 
\begin{eqnarray}
\label{eq:sfr:model:twocomponent}
\dot{\rho} &=&\dot{\rho}_e + \dot{\rho}_s \\
\dot{\rho}_C&=& A_C (t/\tau_C) e^{-t/\tau_C}
\end{eqnarray}
where cosmological time $t$ is measured starting from the beginning of
the universe
and where the two components  decay 
timescales are $\tau_{e,s}=$ 1.5 and 4.5 Gyr, respectively \citep[see Section 2
and Table 2 of ][]{NagamineTwoComponentSFR2006}.
These normalization constants $A_{e,s}= 0.23, 0.15 M_\odot{\rm
  yr}^{-1}{\rm Mpc}^{-3}$ were chosen by \citet{NagamineTwoComponentSFR2006} so the
integrated amount of elliptical and spiral star formation  agrees with
(i) the cosmological baryon census [$\Omega_*\approx 0.0022$; see
\citet{2004ApJ...616..643F,2005RSPTA.363.2693R} and references
therein]; (ii)  the expected degree 
of gas recycling from one generation of stars to the next;  and
(iii) the relative masses in 
different morphological components ($60\%:40\%$).   Explicitly, these
two constants are chosen so
$\int\dot{\rho}_e/\rho_c=\Omega_*/0.55\times0.6$ and $\int
\dot{\rho}_e/\rho_c=\Omega_*/0.55\times0.4$, respectively.

Each component forms stars in its own distinctive conditions, set by
comparison with observations of the Milky Way and elliptical galaxies.
We assume mass converted into stars in the fairly steady ``spiral''
component is done so using 
solar metallicity and with a fixed high-mass IMF power law [$p=-2.7$
in the broken-power-law Kroupa IMF; see
\cite{KroupaClusterAverageIMF2.7}].  
On the other hand, we assume stars born in the   ``elliptical'' component are
drawn from a
broken power law IMF with high-mass index within  $p\in[-2.27,
-2.06]$ and metallicity $Z$ within $0.56<Z/Z_\odot<2$.
These elliptical
birth conditions agree with observations of both old ellipticals in the
local universe \citep[see][and references therein]{astro-ph..0605610} as well as of young starburst clusters
\citep[see][and references therein]{2005ApJ...631L.133F,ADM:Zha99}.

\subsection{Binary pulsar merger rates in the MilkyWay}
\label{sec:obs:rateconstraints}
If binary neutron stars are the source of short GRBs, then the number
of short GRBs seen in spirals should be intimately connected to the number
of  binary pulsars in the Milky Way that are close enough to merge
through the emission of gravitational radiation.
Four unambiguously merging double pulsars have been found within the
Milky Way using pulsar surveys with well-understood selection effects.
\citet{Chunglee-nsns-1} developed a statistical method to estimate the
likelihood of double neutron star formation rate estimates, designed to account
for the small number of known systems and their associated
uncertainties.  
\citet{2004ApJ...614L.137K}
 summarize the latest results of this
analysis: taking into account the latest pulsar data,  a standard
pulsar beaming correction factor $f_b=6$ for the unknown beaming
geometry of PSR~J0737--3037, and a likely distribution of pulsars in the
galaxy (their model 6),  they constrain the
rate to be between $r_{\rm MW} = 16.9 {\rm Myr}^{-1}$ and $292.1 {\rm Myr}^{-1}$.
($95\%$ confidence interval)\footnote{%
The range of binary neutron star merger rates that we expect to
contains the true present-day rate has continued to evolve as our
knowledge about existing binary pulsars and the distribution of
pulsars in the galaxy has changed.  The range quoted here reflects the
recent calculations on binary pulsar merger rates, and corresponds to
the merger rate confidence interval quoted in 
\citet{PSmoreconstraints} (albeit with a different convention for
assigning upper and lower confidence interval boundaries).
In particular, this estimate does not incorportate conjectures regarding a
possibly shorter lifetime of PSR~J0737-3037,   as described in
\citet{Chunglee-nsns-proceedings}.   The properties of this pulsar
effectively determine the present-day merger rate, and small changes
in our understanding of those properties can significantly change the
confidence interval presented.
}.

Assuming all spiral galaxies to form stars similarly to our Milky Way,
then the merger rate density in spirals at present ${\cal
  R}_{\embrace{s}}(T)$ must agree with the product of the formation rate per galaxy $r_{\rm
  MW}$ and the density of spiral galaxies $n_{s}$.  Based on the ratio
of the blue light density of the universe to the blue light
attributable to the Milky Way,  the density of Milky Way-equivalent
galaxies lies between 
$0.75 \times 10^{-2}{\rm Mpc}^{-3}$ and $2\times 10^{-2} {\rm Mpc}^{-3}$
(see  \citet{1991ApJ...380L..17P}, \citet{KNST},
\citet{2004ApJ...612..364N},
\citet{LIGO-Inspiral-s3s4-Galaxies}
 and references therein). 
We therefore expect the merger rate density due to spirals at present
to lie between 
$0.15 {\rm Myr}^{-1}{\rm Mpc}^{-3}$ 
and $5.8 {\rm Myr}^{-1}{\rm Mpc}^{-3}$ (with better than 95\% confidence).

\section{Predictions for Short GRBs}
\label{sec:results:full}
\subsection{Population synthesis simulations}
\label{sec:popsyn}
We study the formation of compact objects with the \emph{StarTrack}
population synthesis code, first developed by  
\citet{StarTrack} and recently significantly extended as described in
detail in \citet{StarTrack2}; see \S 2 of \citet{2006astro.ph.12032B}
for a succinct description of the changes between versions.   

 Since our understanding of the
evolution of single and binary stars is incomplete, this code parameterizes
several critical physical processes with a great many parameters
($\sim 30$), many of which influence compact-object formation
dramatically; this is most typical with all current binary population synthesis codes used by various groups. 
For the \texttt{StarTrack} population synthesis code, in addition to the IMF and
metallicity (which vary depending on whether a binary is born in an
elliptical or spiral galaxy), seven parameters strongly
influence compact object merger rates: the supernova kick distribution
(modeled as the superposition of two
independent Maxwellians, using three parameters: one parameter for the
probability of drawing 
from each Maxwellian, and one to characterize the dispersion of each
Maxwellian), the solar wind strength, the 
common-envelope energy transfer efficiency, the fraction of angular
momentum lost to infinity in phases of non-conservative mass transfer,
and the 
relative distribution of masses in the binary.    
Other parameters,
such as the fraction of stellar systems which are binary (here,
we assume all are, i.e., the binary fraction is equal to $1$) and the distribution of initial binary parameters,
are comparatively well-determined  (see e.g..\cite{1983ARAA..21..343A},
\cite{1991AA...248..485D} and references
therein).\footnote{Particularly for the application at hand -- the
  gravitational-wave-dominated delay between binary birth and merger
  -- the details of the semimajor axis distribution matter little.
  For a similar but more extreme case, see \cite{clusters-2005}.}
%
Even for the less-well-constrained parameters,  some inferences 
have been drawn from previous
studies, either more or less directly (e.g., via observations of
pulsar proper motions, which  presumably relate closely to supernovae
kick strength; see, e.g., \citet{HobbsKicks}, \citet{ArKicks},
\citet{2006ApJ...643..332F} and references therein)
or via comparison of  some subset of  binary population synthesis
results with observations 
(e.g.,  \S 8 of \citet{StarTrack2}, \citet{2006AA...460..209V},
\citet{2005MNRAS.356..753N}, \citet{2002MNRAS.337.1004W},
\citet{2002ApJ...565.1107P} and references therein).
Based on these and other comparisons,  while other parameters entering
into population synthesis models can 
influence their results equally strongly, these particular seven
parameters are the least constrained observationally. 
For this reason, despite observational suggestions that point towards
preferred values 
for these seven parameters -- and despite the good agreement with short
GRB and other observations obtained when using these preferred values
(\citet{ChrisShortGRBs}) -- in this paper we will conservatively examine the
implications of a \emph{plausible range} of each of these parameters.
More specifically, despite the Milky Way-specific studies of
\citet{PSconstraints,PSmoreconstraints} (which apply only to spirals,
not the elliptical galaxies included in this paper), 
in this study we will continue to assume all seven parameters are
unknown, drawn from the plausible parameter ranges described in
\citet{PSmoreconstraints}.


As noted previously in \S~\ref{sec:sfr}, we perform simulations of two
different classes of star-forming conditions: ``spiral'' conditions,
with $Z=Z_\odot$ and a high-mass IMF slope of $p=-2.7$, and
``elliptical'' conditions, with a much flatter IMF slope and a range
of allowed metallicities $0.56<Z/Z_\odot<2$.

\noindent \emph{Archive selection}:  Our collection of population
synthesis simulations consists of roughly $3000$ and $800$ simulations
under spiral and elliptical conditions, respectively.  
Our
archives are highly heterogeneous, with binary sample sizes $N$ that spread
over a large range.\footnote{In practice, the
  sample size is often chosen to insure a fixed number of some type of
event.  As a result, usually the sample size $N$ and the number of
\emph{any} type of event $n$ are correlated.}   A significant fraction 
of the smaller simulations were run with parameters corresponding to
low merger rates, and \emph{have either no BH-NS or no NS-NS merger
  events}.   Therefore, though the set of \emph{all} population
synthesis simulations is unbiased, with each member having randomly
distributed model parameters,
the set of all simulations with one or more events is slightly biased
towards simulations with higher-than-average merger rates.  Further,
the set of simulations with \emph{many} events, whose properties (such
as the merger rate) can be very accurately estimated, can be very
strongly biased towards those models with high merger rates.
Fortunately, as discussed at much greater length in the Appendix,
the set of  simulations with 
$n N \ge \myratio$
and $n>\mynmin$ has small selection bias and  enough simulations 
(976 and 737 simulations NS-NS and BH-NS binaries under spiral-like
conditions, as well as 734 and
650 simulations under elliptical conditions, respectively) to explore
the full range of population synthesis results, while simultaneously
insuring each simulation has enough events to allow us to reliably
extract its results. 

\subsection{Results of simulations}
%
From each population synthesis calculation ($\alpha$) performed under
elliptical or spiral conditions ($C=e,s$) and for each
final result ($K$), we can estimate:
(i) the number of final $K$ events per unit mass of binary mergers progenitors, i.e., the
\emph{mass efficiency} ($\lambda_{C,\alpha,K}$); and
(ii) the probability $P_{c,\alpha,K}(<t)$ that given a progenitor of
$K$ the event $K$ (e.g., a  BH-BH merger) occurs by time $t$ since the
formation of $K$.
Roughly speaking, for each simulation we take the observed sample of  $n$
binary progenitors of $K$, with $M_{1\ldots n}$ and delay times
$t_{1\ldots n}$, and estimate
\begin{eqnarray}
\label{eq:popsyn:lambda}
\lambda &=& 
\frac{n}{N} \frac{f_{cut}}{\left< M \right>} \\
\label{eq:popsyn:Pdelay}
P_m(<t)&=&  \sum_j \Theta(t-t_j)
\end{eqnarray}
where $\Theta(x)$ is a unit step function;  $N$ is the total number of
binaries simulated, from which the $n$ progenitors of $K$ were drawn;
$\left<M\right>$ is the average mass of all possible binary progenitors; and 
$f_{cut}$ is a correction factor accounting for the great many very
low mass binaries (i.e., with primary mass $m_1<m_c=4 M_\odot$) not
included in our simulations at all.  Expressions for both
$\left<M\right>$ and $f_{cut}$ in terms of population synthesis model
parameters  are provided in Eqs. (1-2) of \citet{PSutil2}.  In practice, $P_m(t)$ and $dP_m/dt$ are estimated with smoothing kernels, as
discussed in  Appendix \ref{ap:smoothingSystematics}.
Given the characteristic sample sizes involved (e.g., $n>200$ for NS-NS), we expect  $P_m$ to   have absolute
error less than 0.05 at each point (95\% confidence) and $dP_m/dt$ to
have rms relative error less than 20\%  (95\% confidence).  Since these errors
are very small in comparison to uncertainties in other quantities in
our final calculations
(e.g.,  the star
formation rate of the universe), we henceforth
ignore errors in  $P$ and
$dP/dt$.

\begin{figure}
\includegraphics[width=\columnwidth]{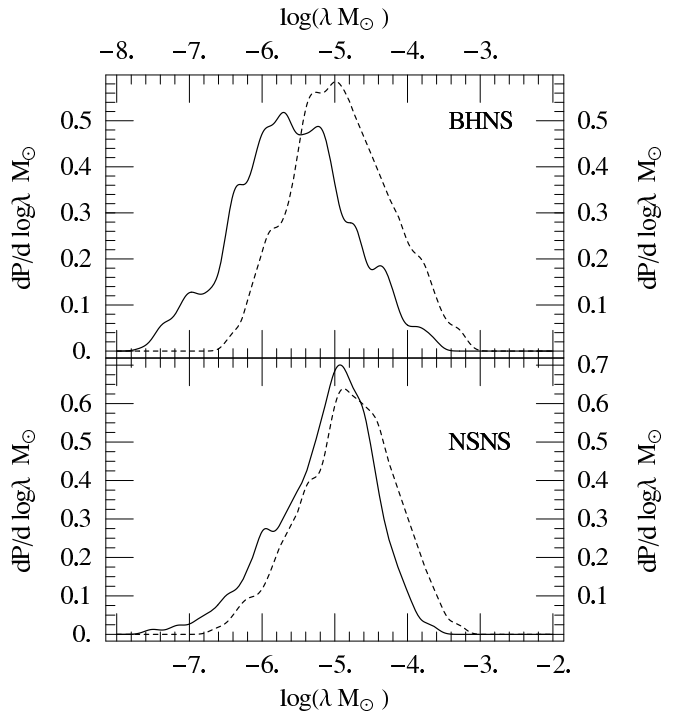}
\caption{\label{fig:popsyn:distrib:epsilonlambda} Smoothed histograms
  of the 
  mass efficiency $\lambda$   [Eq.~(\ref{eq:popsyn:lambda})] of
  simulations used in our calculations, shown for spiral (solid) and
  elliptical (dotted) birth conditions.  
As expected given the
differences in the IMF,
elliptical galaxies produce BH-NS binaries roughly three times more
efficiently than spirals.  However, apparently because our population
synthesis sample involves highly correlated choices for $n$ and $N$
(see the Appendix and Figure \ref{fig:archiveselect:scatter}), our
distribution of NS-NS mass  efficiencies remains biased,
producing identical distributions for both elliptical and spiral birth conditions.
}
\end{figure}

\begin{figure*}
\begin{centering}
\includegraphics[width=0.8\textwidth]{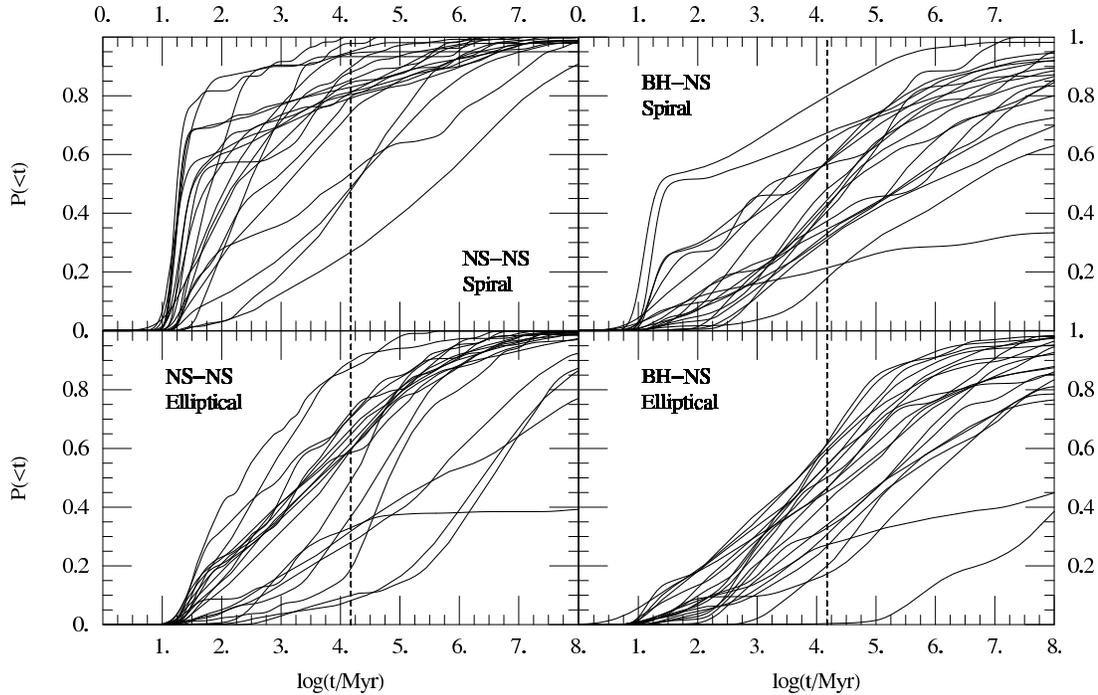}
\end{centering}
\caption{\label{fig:popsyn:distrib:pcum} Cumulative probabilities
  $P_m(<t)$ that a NS-NS binary (left panels) or BH-NS binary (right panels) will merge in time less than $t$, for twenty \emph{randomly-chosen} population synthesis models,
  given spiral (top) and elliptical (bottom) star forming conditions.
  A vertical dashed line indicates the age of the universe.
  For the sample sizes involved, these distributions are on average
  accurate to within 0.05 everywhere (with 90\% probability); see
  Figure \ref{fig:fits:PcumSmoothingWorks}.  
}
\end{figure*}

\begin{figure*}
\begin{centering}
\includegraphics[width=\columnwidth]{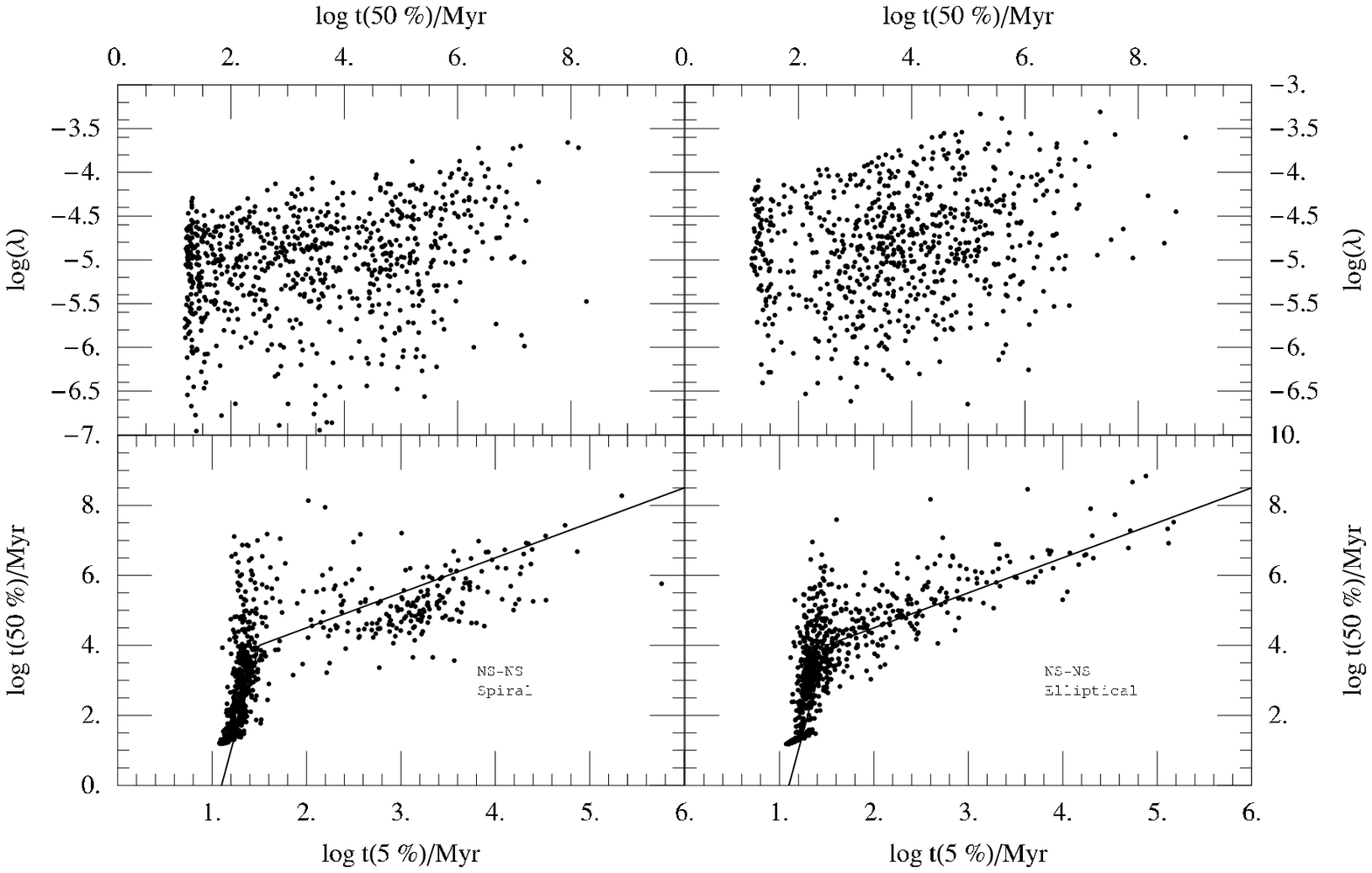}
\includegraphics[width=\columnwidth]{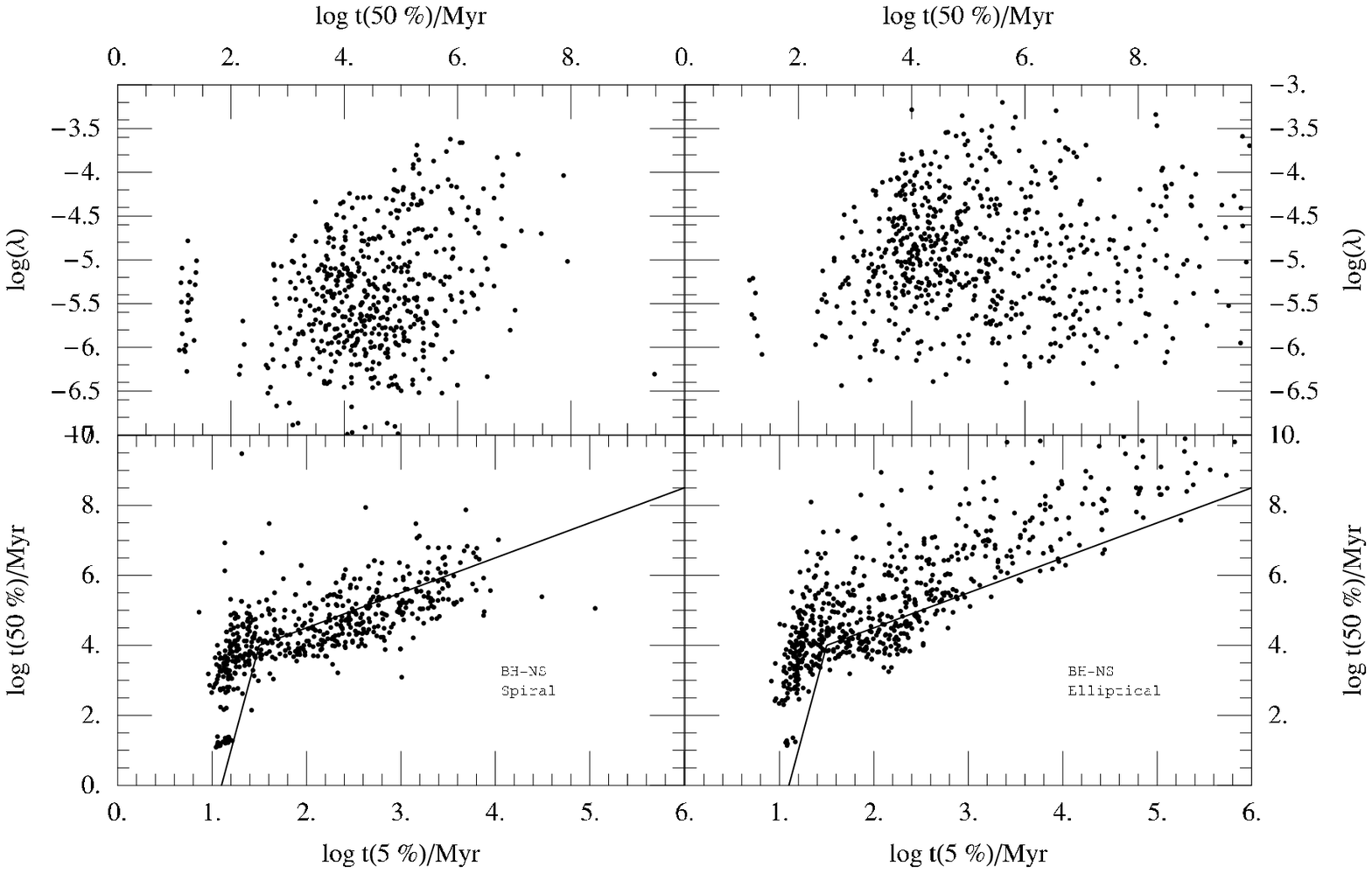}
\end{centering}
\caption{\label{fig:popsyn:distrib:correlations}Scatter plots to
  indicate correlations between results of various simulations.  Top panels:
  Scatter plot of mass efficiency $\log \lambda$ and average delay
  time $\log t(50\%)$.  Bottom panels: Scatter plot of $\log t(5\%)$ versus
  $\log t(50\%)$; also shown is the analytic estimate of 
  Eq. (\ref{eq:popsyn:results:semianalyticCorrelationT}).
Left panels
  indicate spiral conditions; right panels indicate elliptical
  conditions.  
  Despite the differences between these two types of simulations
 (i.e., metallicity and initial mass function), the range of delay
 time distributions and mass efficiencies are  largely
 similar (i.e., the corresponding left and right panels resemble one another).
}
\end{figure*}

Figures \ref{fig:popsyn:distrib:epsilonlambda},
\ref{fig:popsyn:distrib:pcum}, and
\ref{fig:popsyn:distrib:correlations} show explicit results drawn from
these calculations.  From these figures, we draw the following
conclusions:

\noindent \emph{Uncertainties in binary evolution significantly affect
  results}: As clearly seen by the range of possiblities allowed in
Figures \ref{fig:popsyn:distrib:epsilonlambda} and
\ref{fig:popsyn:distrib:pcum},
our imperfect understanding of binary evolution implies we must permit
and consider models with a wide range of mass efficiencies $\lambda$
and delay time distributions $P_m(<t)$.

\noindent\emph{The merger time distribution is often well-approximated
  with a one parameter family of distributions, $dP_m/dt\propto 1/t$}:
As suggested by the abundance of near-linear distributions in Figure
\ref{fig:popsyn:distrib:pcum}, the delay time distribution $P_m$ is
almost always  linear $\log t$.  Further, from the relative
infrequency of curve crossings, the \emph{slope} of $P_m$ versus $\log
t$ seems nearly constant.  As shown in the bottom panels of Figure
\ref{fig:popsyn:distrib:correlations},  this constant slope shows up as a strong
correlation between the times $t(5\%)$ and $t(50\%)$ at which $P_m$
reaches 0.05 and 0.5 when $\log t(5\%)/Myr>1.5$: 
\begin{equation}
\label{eq:popsyn:results:semianalyticCorrelationT}
\log t(50\%) \approx \left\{
\begin{array}{ll}
     \log t(5\%) + 2.5 & \text{if } \log t(5\%)>1.5 \\
     10 \log t(5\%) -11 & \text{if } \log t(5\%)<1.5
\end{array}
\right .
\end{equation}

\noindent \emph{The merger time distribution is at most weakly
  correlated with the mass efficiency}: Finally, as seen in the top
panels of Figure \ref{fig:popsyn:distrib:correlations}, a wide range of
efficiencies are consistent with each delay time distribution.  The
maximum and minmum mass efficiency permitted  increase marginally with
longer median delay times $t(50\%)$ -- roughly an order of magnitude
over five orders of magnitude of $t(50\%)$.  But to a good
approximation, the mass efficiencies and delay times seem to be uncorrelated.

\subsection{Converting simulations to predictions}
Each combination of  heterogeneous population synthesis model, star formation
history, and source model leads to a specific physical signature, embedded in
observables such as the relative
frequencies with which short GRBs appear in elliptical and spirals, the
average age of progenitors in each component, and the
observed distribution of sources' redshifts and luminosities.  All of
these quantities, in turn, follow from the two functions ${\cal R}_C(t)$,
the merger rate  per unit comoving volume in $C=$ elliptical or spiral galaxies.

The rate of merger events per unit comoving volume
is uniquely determined from (i) the SFR of the components of the
universe $d\rho_{\{C\}}/dt$; (ii) the mass efficiency $\lambda_{\embrace{C}}$ at
which $K$ mergers occur in each component $C$; and (iii) the
probability distribution $dP_{m\{C\}}/dt$ for merger events to occur
after a delay time $t$ after star formation:
\begin{eqnarray}
\label{eq:comovingRateNet}
{\cal R}(t) &=& \sum_{C}{\cal R}_{\embrace{C}} \\
\label{eq:int:comovingRate}
{\cal R}_{\{C\}}(t) &=& \int dt_b \lambda_{\{C,K\}}\frac{dP_{m\{C\}}}{dt}(t-t_b) \frac{d\rho_{\{C\}}}{dt}(t_b)
\end{eqnarray}
Though usually ${\cal R}_{\embrace{C}}(t)$ is experimentally
inaccessible, because our source and detection models treat elliptical
and spiral hosts identically,\footnote{In practice, gas-poor
  elliptical hosts should produce weaker afterglows.  Since
  afterglows are essential for host identification and redshifts,
  elliptical hosts may be under-represented in the observed sample.} 
the ratio uniquely determines the fraction $f_{s}$ of spiral hosts at a given
redshift: 
\begin{equation}
\label{eq:spiralFraction}
f_{s}(z) = {\cal R}_{\embrace{s}}/({\cal R}_{\embrace{e}}+{\cal R}_{\embrace{s}})
\end{equation}
Additionally, 
as described in \S\ref{sec:obs:rateconstraints}
observations of NS-NS in the Milky Way constrain the
present-day merger rate of short GRB progenitors
(${\cal  R}_{\embrace{s}}(T_{\rm universe})$), \emph{if} those
progenitors are double neutron star binaries.

Unfortunately, the relatively well-understood \emph{physical} merger rate 
distributions $R_{\embrace{C}}$ are  disguised by strong
observational selection effects described in
\S~\ref{sec:grb}, notably in the short GRB luminosity function.
Based on our canonical  source model, we predict the detection rate
$R_{D}$ of short GRBs from
to be given by 
\begin{eqnarray}
\label{eq:rateCosmologyIntegral}
R_{D} &=&\sum_{C}R_{D \embrace{C}} \\
R_{D \embrace{C}} &=& f_d^{-1}\int  {\cal R}_{\embrace{C}} P_\embrace{C}(z) 4 \pi r^2 c dt \\
   &\approx &  
   \frac{\dot{N}_{min }}{f_d f_b F_d} 
   \int  c dt\,
     \frac{{\cal R}_{\embrace{C}}(t)}{(1+z)k(z)} \nonumber
\end{eqnarray}
where the latter approximation holds for reasonable 
$\dot{N}_{min}/f_b<10^{57} {\rm s}^{-1}$  (i.e., below values corresponding to
observed short bursts).
While the detection rate depends sensitively on the source and
detector models, within the context of our source model the differential
redshift 
distribution $p(z)$
\begin{eqnarray}
\label{eq:redshiftDistrib}
p(z) dz &\propto &
     \frac{dt}{dz}
 \sum_{C,K}\frac{{\cal R}_{\embrace{C}}(t(z))}{(1+z)k(z)}\frac{\dot{N}}{f_{b}}
\end{eqnarray}
and the cumulative redshift distribution $P(<z)\equiv \int_0^z p(z)dz$
do not depend on the source or detector model  \citep{Nakar}.

\noindent \emph{Detected luminosity distribution}:  In order to be
self-consistent, the predicted luminosity distribution should agree
with the observed peak-flux distribution seen by BATSE.  However,
so long as
$\dot{N}_{min}$ is small for all populations,  our hypothesized power-law form  $\dot{N}^{-2}$ for the GRB
luminosity function 
 immediately implies the detected cumulative peak flux distribution
$P(>F)$ is precisely consistent: $P(>F)=(F_d/F)$, independent of
source population; see for example the discussion in \citet{Nakar}
near their Eq. (5).    While more general source population models
produce observed luminosity functions that contain some information about
the redshift distribution of bursts -- for example,
\cite{GuettaPiran2006} and references therein employ broken power-law
luminosity distributions;
alternatively, models could introduce correlations between brightness
and spectrum
-- so long as most sources remain unresolved (i.e., small
$\dot{N}_{\rm min}$),  the observed peak flux distribution largely
reflects the intrinsic brightness distribution of sources.   
Since \citet{Nakar} demonstrated this particular brightness distribution
corresponds to  the observed BATSE flux distribution,  we learn
nothing new by comparing the observed peak flux distribution with
observations and henceforth omit it.

\subsection{Predictions for short GRBs}
\label{sec:results:general}

Given \emph{two} sets of  population synthesis model parameters --
each of which 
describe star formation in elliptical and spiral galaxies,
respectively --  the tools
described above provide a way to extract merger and GRB detection
rates, \emph{assuming all BH-NS or all NS-NS mergers produce} (possibly
undetected) \emph{short GRB events}.  
Rather than explore the  (fifteen-dimensional: seven parameters for
spirals and eight, including metallicity, for ellipticals) model space
exhaustively, we explore a limited array\footnote{We draw our
  two-simulation ``model universe'' from two collections of order
  800 simulations that satisfy the constraints described in the
  Appendix, one for ellipticals and one for spirals.
  Computational inefficiencies in our postprocessing pipeline 
  prevent us from performing   a thorough survey of all $\sim 10^5$
  possible combinations of existing spiral and elliptical simulations. 
} 
of 500 ``model universes''
 by  (i) 
\emph{randomly}\footnote{At present, our population synthesis archives
for elliptical and spiral populations were largely distributed
independently; we cannot choose pairs of models with \emph{similar or
  identical} parameters for, say, supernovae kick strength
distributions.  The results of binary evolution from elliptical and
spiral star forming conditions, if anything, could be substantially
more correlated than we can presently model. We note however that
there is no a priori expectation nor evidence that evolutionary
parameters should indeed be similar in elliptical and spiral
galaxies.}  
selecting  two population synthesis simulations, one each associated with
elliptical ($e$) and spiral ($s$) conditions;  (ii) estimating for
each simulation the mass efficiency ($\lambda_{e,s}$) and delay
time distributions ($P_{e,s}$);  (iii)
constructing the net merger rate distribution ${\cal R}$ using
Eqs. (\ref{eq:comovingRateNet},\ref{eq:int:comovingRate});  and (iv)
integrating out the associated expected redshift distribution $p(z)$  [Eq.~(\ref{eq:redshiftDistrib})].

\begin{figure}
\includegraphics[width=\columnwidth]{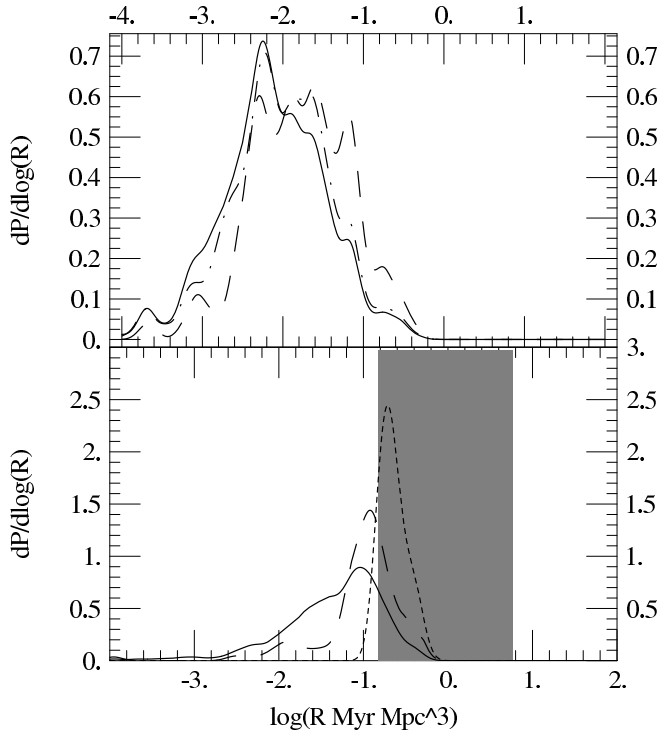}
\caption{\label{fig:results:MergerRates} 
The distribution of merger rate densities in spiral-type galaxies
${\cal R}_s$ for BH-NS mergers (top) and NS-NS mergers (bottom); the
solid curve includes all simulations,  the dashed
curve only those simulations reproducing the observed redshift distribution;
and the dotted curve (bottom panel only) only those simulations reproducing the NS-NS
merger rate in spiral galaxies derived from an analysis of binary
pulsars (also shown on the bottom panel, in
gray; see  \S\ref{sec:obs:rateconstraints} for details); and the dot-dashed
curve (top panel only) includes only those simulations which, under
the most optimistic assumptions, predict  short GRBs
should occur at least as frequently as has been seen.  The bottom panel
in particular should be compared with  Figure 3 (top panel)
of \citet{PSmoreconstraints}.
}
\end{figure}

\begin{figure}
\includegraphics[width=\columnwidth]{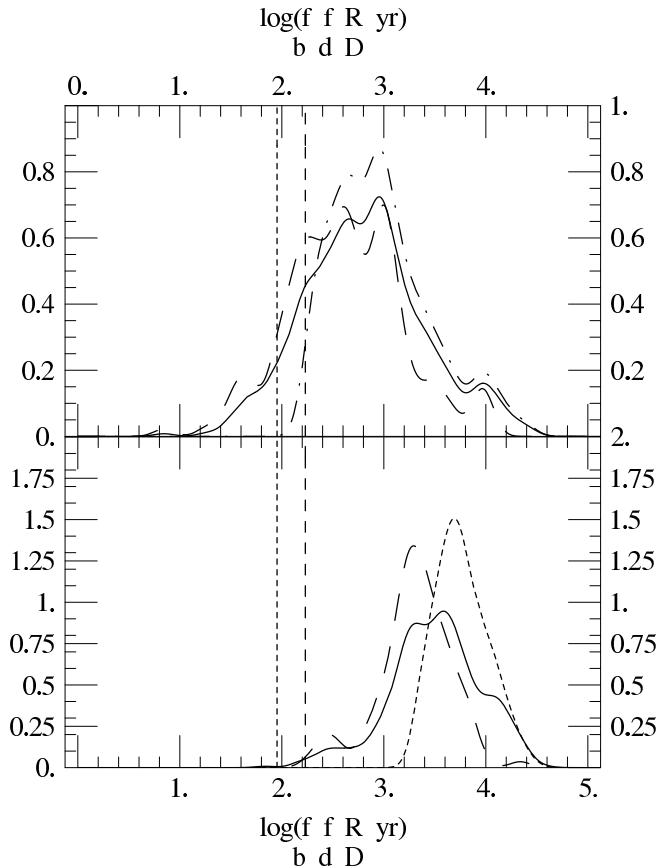}
\caption{\label{fig:results:DetectionRatesModFaintestBurst} 
Distribution of predicted \emph{all-sky, beaming-corrected} short GRB detection rates  $\log R_D f_b f_d$, if bursts arise from BH-NS (top)
and NS-NS (bottom) mergers  \emph{and} all bursts produce a \emph{higher} peak
flux of 50-300 keV photons than the least-luminous burst in our sample (with
$\dot{N}\simeq 3\times 10^{55} \unit{s}^{-1}$), compared to the
observed BATSE (dashed vertical line) and Swift (dotted) all-sky
detection rate.  
In other words,
a plot of the most optimistic predictions of each two-component
``model universe'' for the  burst detection rate: assuming a detector has all sky visibility, that all
burst emit isotropically with no beaming, and that no bursts are faint
enough to be missed. 
 Almost all  models
produce at least as many bursts as are observed; all models can
therefore reproduce observations by choosing  some low minimum peak photon
luminosity $\dot{N}_{\rm min}$, typically $10^2$-$10^3$ times lower
than the one of  the faintest burst seen.  
As in Figure \ref{fig:results:MergerRates}, the solid curves include
all 500 ``model universes''; the dashed curves include only those
``model universes'' which reproduce the short GRB redshift
distribution; the dotted curve (bottom panel only) includes only models
which are reasonably consistent with the present-day double neutron star merger
rate in the Milky Way; and the dot-dashed curve (top panel only)
includes only those simulations which, under
the most optimistic assumptions, predict  short GRBs
should occur at least as frequently as has been seen.
}
\end{figure}

\begin{figure*}
\includegraphics[width=0.9\textwidth]{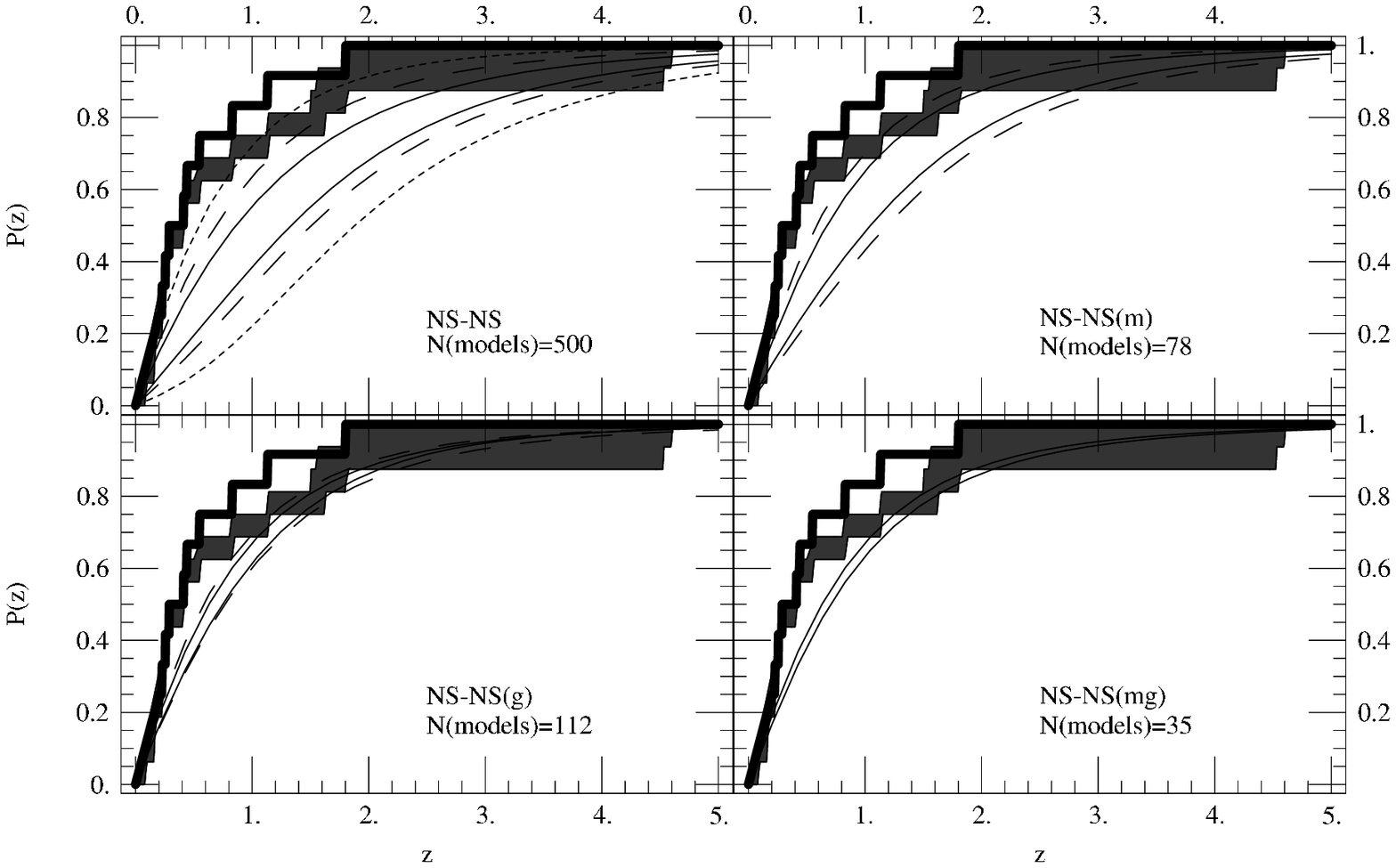}
\caption{\label{fig:results:NSNS}   
Demonstration that population
  synthesis models can reproduce short GRB redshift
  distributions, assuming NS-NS mergers produce all short GRBs.  As in
  Figure \ref{fig:results:BHNS}, the top right and left panels compare
  the range of short GRB redshift distributions expected from our
  two-component model : 1\%/99\% (dotted), 10\%/90\% (dashed), and
  25\%/75\% (solid) redshift distributions are overlaid on the
  observed cumulative redshift distribution, both for all simulated
  models (top left) and those models which remain everywhere close to
  the observed redshift distribution (bottom left).
  Top right: As above, but including only those NS-NS models which
  produce a NS-NS merger rate in agreement with observations of the
  Milky Way merger rate (Figure \ref{fig:results:MergerRates}).  
  Bottom right: As above, but including only those NS-NS models which
  reproduce the number of binary pulsars seen in the Milky Way and the
  short GRB redshift distribution.
}
\end{figure*}

\begin{figure*}
\includegraphics[width=0.9\textwidth]{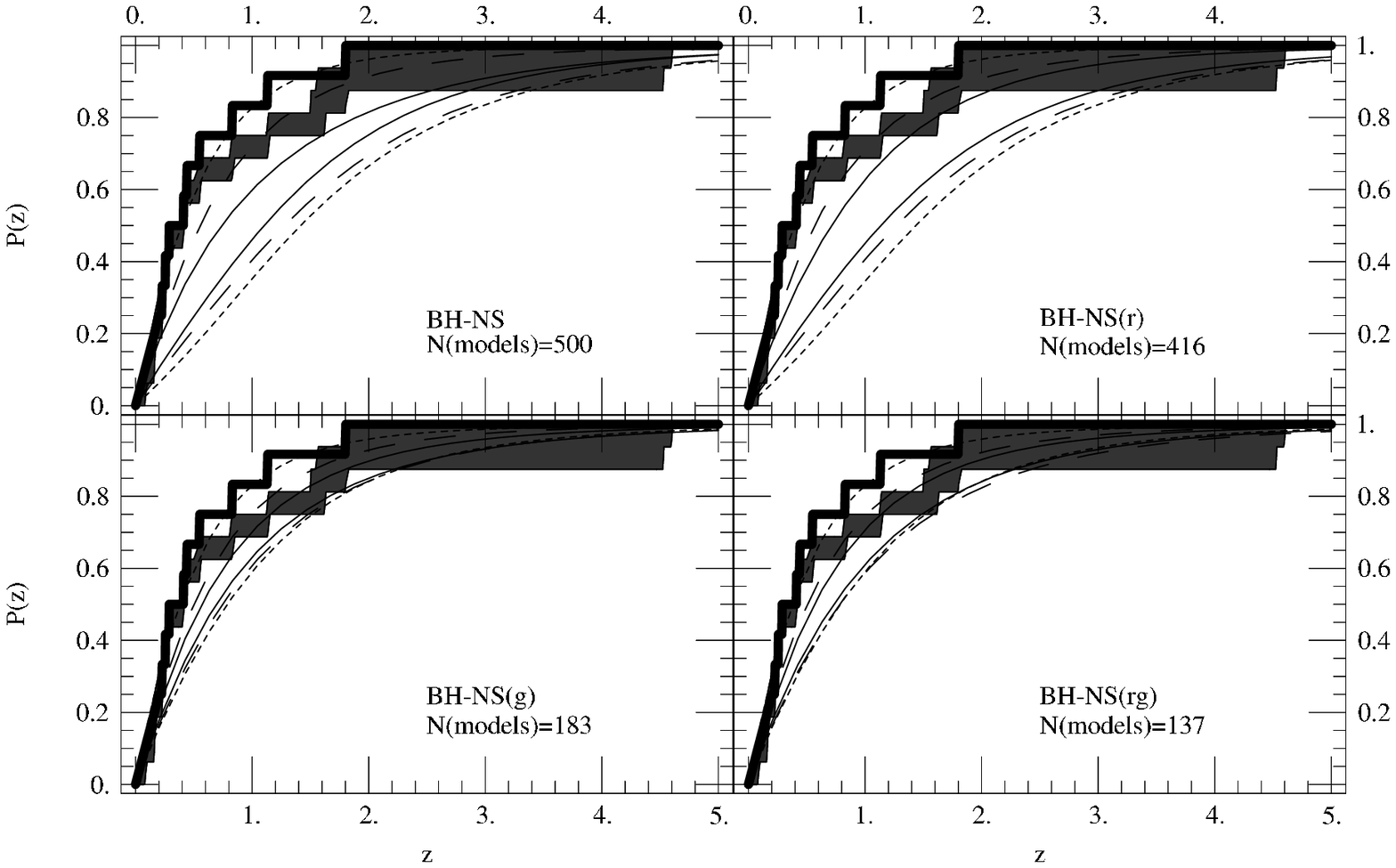}
\caption{\label{fig:results:BHNS}  
Demonstration that population
  synthesis models can reproduce short GRB redshift
  distributions, assuming BH-NS mergers produce all short GRBs.  The
  jagged curve and shaded regions provide the cumulative redshift
  distribution for observed short GRBs (Fig. \ref{fig:data:grbs}).
  Top left panel: The smooth curves illustrate the range of short GRB redshift
  distributions; out of the 500 simulated, the two solid curves
  correspond to the curves with the 25th and 75th percentile values of
  $P(0.1)$; the dashed curves to the 10th and 90th percentiles; and
  the dotted curves to the first and 99th percentiles.
  Bottom left panel: As above, but including only those models which produce
  redshift distributions which differ from observations by less than
  $0.38$ (i.e., a 5\% Kolmogorov-Smirnov false alarm probability).
  Top right panel: As previously, but including only those model
  which, when given the most favorable assumptions, still predict too
  few short GRB detections.
  Bottom right panel: As previously, but requiring both the predicted short GRB
  rate and redshift distribution be simultaneously consistent with GRB
  observations.
}
\end{figure*}

\begin{figure}
\includegraphics[width=\columnwidth]{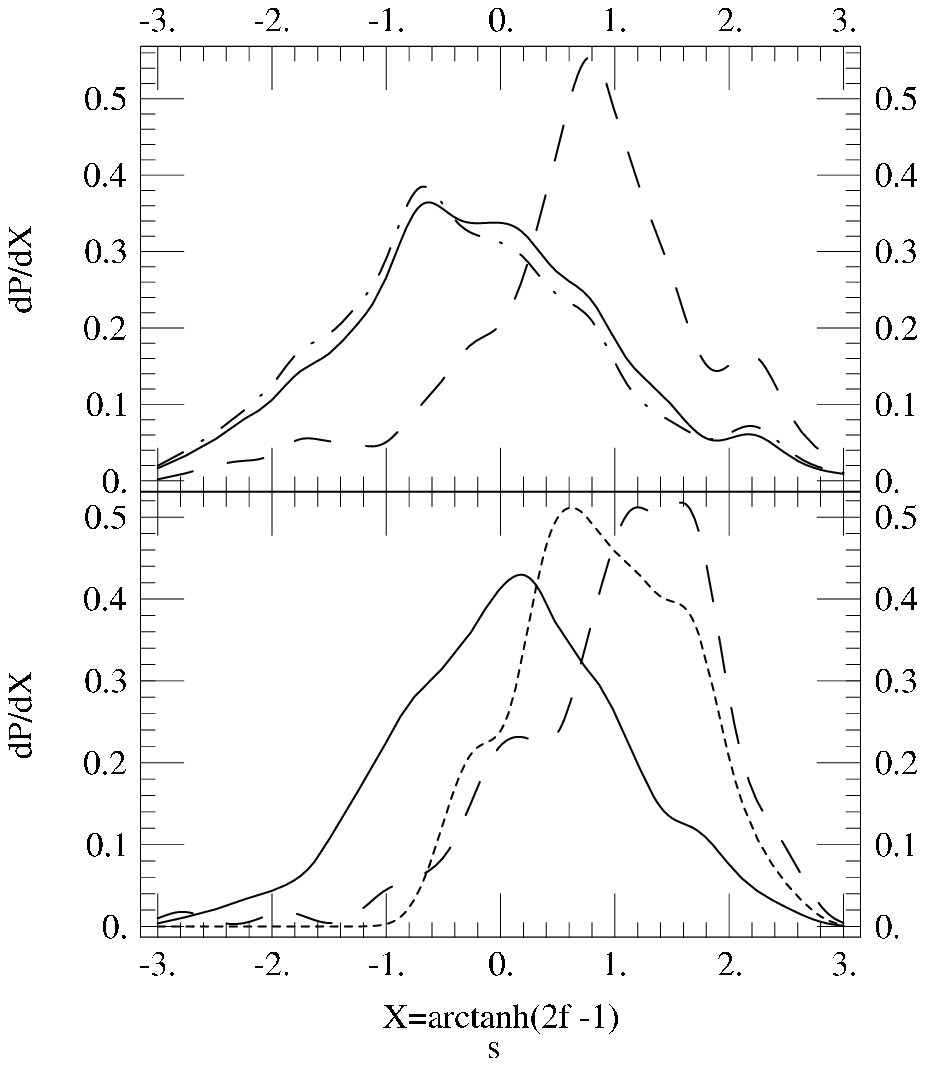}
\caption{\label{fig:results:SpiralFraction} 
Distribution of the fraction of BH-NS mergers (top panel) and NS-NS mergers (bottom panel)
expected to occur in spiral galaxies \emph{at the present day}
[Eq.~(\ref{eq:spiralFraction}) evaluated at $z=0$]. Solid curves
include all models; dashed lines include only those models reproducing
the redshift distribution; dotted line includes only those models
reproducing the merger rate of NS-NS binaries in spiral galaxies,
based on the Milky Way; and the dot-dashed line includes only models
which could possibly produce as many short GRB events as are observed.
 To better indicate a very high or very 
  low fraction $f_{s}$ of mergers occurring in spirals, we plot the
  distributions above versus $X=\tanh^{-1}(2f_{s}-1)$.
}
\end{figure}

The results of these calculations are presented in Figures
\ref{fig:results:MergerRates},
\ref{fig:results:DetectionRatesModFaintestBurst},
 \ref{fig:results:NSNS}, \ref{fig:results:BHNS},
and \ref{fig:results:SpiralFraction}.  [These figures also compare our calculations' range of results to
observations of short 
GRBs (summarized in Table \ref{tab:grbs}) and merging Milky Way binary
pulsars \citep{Chunglee-nsns-1}; these comparisons will be discussed
extensively in the next section.]  More specifically, these five
figures illustrate the distribution of the following four quantities that we extract from
each ``model universe'':

\noindent \emph{Binary merger rates in present-day spiral galaxies}:   To enable a clear
comparison between our multi-component calculations, which include
both spiral and elliptical galaxies, and other
merger rate estimates that  incorporate only present-day star
formation in Milky Way-like galaxies, 
the two solid curves in Figure \ref{fig:results:MergerRates}  show the
distributions  of
present-day NS-NS and BH-NS merger rates in \emph{spiral} galaxies
seen in the respective sets of 500 simulations.  

In principle, the BH-NS and NS-NS merger rates should be weakly
correlated, as the processes (e.g., common envelope) which drive double neutron star to merge
also act  on low-mass BH-NS binaries, albeit not always similarly; as
a trivial example, mass transfer processes that force binaries together more
efficiently may  deplete the population of NS-NS binaries in earlier
evolutionary  phases while simultaneously bringing distant BH-NS
binaries close enough to merge through gravitational radiation.
Thus, a simulation which contains enough  BH-NS
binaries for us to estimate  its delay time
distribution $dP/dt$ need not have produced similarly many NS-NS binaries.
For this reason, we constructed \emph{two independent sets} of 500 ``model
universes'', one each for BH-NS or NS-NS models.
However, as a corollary, the randomly-chosen simulations used to
construct any given BH-NS ``model universe'' 
 need not have enough  merging NS-NS to
enable us to calculate the present-day merger rate, and vice-versa.
In particular,  we \emph{never} calculate the double neutron star merger
rates in the BH-NS model universe.    Thus, though the BH-NS and NS-NS
merger rates should exhibit some  correlation, we do not explore it
here.  In particular, in the next section where we compare predictions
against observations, we do not require  the BH-NS
``model universes'' reproduce the present-day NS-NS merger rate.

%
\noindent \emph{Short GRB detection rate}:  As described in detail in 
 \S~\ref{sec:grb},  the fraction of short GRBs
on our past light cone which are \emph{not} seen depends strongly but
simply on
unknown factors such as the fraction of bursts pointing towards us,
which we characterize by $1/f_b$ where $f_b$ is the beaming factor,
and the fraction of bursts luminous enough to be seen at a given
distance, which we characterize by $P(>N_d)$ where $N_d =4\pi r^2
k(z)(1+z) F_d$ is the minimum photon luminosity visible at redshift
$z$.    The short GRB \emph{detection} rate also depends on the
\emph{detector}, including the fraction of sky it covers ($1/f_d$) and
of course the minimum flux $F_d$ to which each detector is sensitive.
To remove these significant ambiguities, in Figure
\ref{fig:results:DetectionRatesModFaintestBurst} we use solid curves
to plot the
distribution of detection rates found for each of our 500 ``model
universes'' (top panel and bottom panel correspond to the BH-NS and
NS-NS model universes, respectively), assuming (i) that no burst is
less luminous than the least luminous burst seen, 
namely, GRB 060505, with an apparent (band-limited) isotropic
luminosity 
$\dot{N}_{\rm min\, seen}\simeq 3\times10^{55}\unit{s}^{-1}$, or $L\simeq 7\times 10^{48}
\unit{erg}\unit{s}^{-1}$ (see Table
\ref{tab:grbs}); (ii) that beaming has been ``corrected'', 
effectively corresponding to assuming isotropic detector
sensitivity and source emission; and (iii) that the detector has a peak flux
detection threshold of $F_d=1\unit{cm}^{-2}\unit{s}^{-1}$,
corresponding roughly to the true BATSE and Swift peak flux thresholds
presented in \S~\ref{sec:grb}.  These choices are conservative;
therefore, our rate estimate for each model universe is an
\emph{upper bound}.


\noindent \emph{Short GRB redshift distribution}:  The short GRB
detection rate depends on the ability of gamma-ray detectors
to distinguish burst events from background noise  and the GRB
luminosity function.   Its selection effects
therefore depend only on the properties of gamma-ray telescopes.  On
the other hand, the short GRB redshift distribution implicitly
contains several other selection effects which have not been factored
into our analysis (e.g., regarding the availability and reliability
of an association  of each burst to a host and the ability of optical
telescope to extract a spectrum and redshift of that host).  Lacking
the ability to characterize these selection effects, we perform the
most straightforward prediction for the distribution of short GRB
redshifts: using  Eq. (\ref{eq:redshiftDistrib}), we assume
all detected short GRB events  have accurate and unambiguous optical
follow-up and redshifts.   Based upon that assumption, we generate two
sets of 500 candidate redshift distributions for short GRBs, assuming
either (i)  the set of merging BH-NS binaries or (ii) the set of  merging NS-NS
binaries is identical to (iii) the set of all short GRBs.
The results are shown in the top left panels of Figures
\ref{fig:results:NSNS} (for NS-NS binaries) and
\ref{fig:results:BHNS} (for BH-NS binaries).
Rather than provide all 500 cumulative redshift distributions, we have sorted
these distributions by their value at $z=0.1$ and plotted indicative curves at chosen percentiles: for example, the dotted curves show
the 50th and 450th cumulative redshift distributions we obtained,
corresponding to the 10th and 90th percentile.  
Because of the limited number of ``model universes'' presently
available to us, we do not presently attempt to resolve the tails of
the distribution; however the range of potential cumulative redshift
distributions includes a low a priori probability tail with members
which are slightly more biased  towards 
low redshift than the cumulative distributions those shown here.


\noindent \emph{Fraction of mergers in spiral galaxies}:  Finally,  in
Figure \ref{fig:results:SpiralFraction} (solid curves) we show the a
priori probability that a fraction $f_s$ of binary 
mergers occur in spiral galaxies, using the scaled variable
$X=\text{arctanh}(2f_s-1)$.  This representation allows us to better illustrate
relative probabilities of  spiral 
fractions $f_s$ that are very near $1$ or $0$; to give a sense of scale, $X=1
(2)$ corresponds to $88\%$ ($98\%$) of all mergers occurring in spiral 
galaxies.   

As seen in Figure \ref{fig:results:SpiralFraction}, a priori we cannot
say whether elliptical or spiral galaxies should host most
presently-occurring binary mergers.  Furthermore, a significant
fraction of models have $|X|>1$ (i.e., in those models, almost all
mergers occur in spirals or ellipticals).  This extreme range for
$f_s$ can be traced back directly to the large range of mass
efficiencies shown in Figure 
\ref{fig:popsyn:distrib:epsilonlambda}.  For a randomly
chosen pair of population synthesis model for elliptical and spiral
galaxies, one will often be significantly larger than the other,
leading to  a spiral fraction $f_s$ near its limits (i.e., $f_s\simeq 0,1$).

\section{Testing Predictions Against Observations}
\label{sec:tests}
To summarize, in this paper we are trying to examine whether either of two very simple
hypotheses are consistent with what we have heretofore seen of short
GRBs and binary mergers are consistent with our understanding of the star
formation history of the universe and of binary stellar evolution. The
two simplest hypothesis are, on the one hand,  that every short GRB is 
a binary BH-NS merger and that every BH-NS
merger produces a short GRB; and on the other hand the corresponding
statement for NS-NS binaries.\footnote{%
We start with the simplest and most tractable pair of hypotheses.  A
more realistic model would allow for a fraction of both BH-NS and
NS-NS mergers to produce short GRBs, along with some proportion of
young neutron stars (in  bursts from soft gamma repeaters, commonly
abbreviated SGRs).  However, with 
so many degrees of freedom, such a model would be very difficult to
constrain without additional observational inputs (e.g., direct
confirmation of the fraction of mergers which are SGRs in the nearby
universe; a reliable measure of the fraction of mergers that occur in
elliptical and spiral galaxies; gravitational wave detection of merger
events; etc.).
}
Further, assuming that in both cases some fraction of all possible
models are consistent with observations, we want to understand the
properties of those consistent models and the physical processes that
require those properties to be as they are.
Of course, these comparisons are affected by observational selection effects.   Notably, \citet{berger-manyfainthosts2006} has suggested
that  the lack of  deep optical follow-up on faint
short GRBs has biased the redshift distribution to low redshift.
Lacking any ability to control an unknown  bias, we proceed with the simplest
possible test: we  compare existing results to
currently known observations.  
 Specifically, we require our
predictions for the binary neutron star merger rate (Figure
\ref{fig:results:MergerRates}, available only for binary neutron star
short GRB ``model universes''), for the short GRB detection rate 
(Figure \ref{fig:results:DetectionRatesModFaintestBurst}), 
and for the short GRB redshift distribution 
(Figures \ref{fig:results:NSNS} and \ref{fig:results:BHNS})  be
statistically consistent with the corresponding observations, all of
which are shown in their respective plots.
We then compare the distribution of \emph{constrained} quantities (in
the aforementioned Figures and in Figures
\ref{fig:results:SpiralFraction} and
\ref{fig:results:EllipticalDelayContours})
with their initial distributions, to understand the physics implied by
observations.\footnote{Because of the limited number of
  constrained models and the large number of underlying parameters
  (fifteen), we do \emph{not} attempt to characterize the underlying
  physical parameters of constraint-satisfying models.  Instead, we
  focus on describing the essential features of models, such as a long
  characteristic delay time or high mass efficiency, that allow a
  model to satisfy observational constraints.}

\subsection{Comparisons I: NS-NS as burst source}
No definitive evidence exists to determine the fraction of
short bursts that are less luminous than the faintest currently known;
to discover the degree to which, if any, short GRBs are beamed\footnote{Evidence has been presented to
  suggest that short GRB emission has a break in one or more
  frequencies, a result that has been interpreted as a ``jet break''
  produced by a beamed jet
\citep[see,e.g.,][and references therein]{2006ApJ...650..261S,2006ApJ...653..462G,NakarReviewArticle2006}.  At present we choose to remain conservative
  regarding beaming until several multi-band observations confirm these
  breaks exist.};
and to demonstrate that all mergers must produce bursts.  Therefore,
we must take the bottom panel of   
Figure \ref{fig:results:DetectionRatesModFaintestBurst} at face value:
\emph{if}  no bursts are less
luminous than those seen, every short GRB model based on NS-NS produces many more
short GRBs than are seen, and therefore every double neutron star
``model universe'' can reproduce the present-day short GRB detection
rate by a proper choice of, for example, the minimum luminosity of
short GRBs.  Therefore, only \emph{two} of the three available observations -- the short GRB redshift
distribution and the set of merging Milky Way binary pulsars --
can reject some of the 500 double neutron star ``model universes.''

The statistics of this comparison are summarized in Figure \ref{fig:results:NSNS}.
Relatively few of our
double neutron star models are consistent with observations of merging
double pulsars in the 
Milky Way\footnote{Fewer still would be consistent should we require agreement with more
binary pulsar rate constraints, as has been shown in \citet{PSconstraints} and
\citet{PSmoreconstraints}. 
}: 78 out of 500 models, or $\simeq 16\%$; see the
top right panel of Figure \ref{fig:results:NSNS}
as well as Figure
\ref{fig:results:MergerRates}.
Among the few models which agree with Milky Way observations, however,
most   have a cumulative redshift distribution within 0.375 of our
family of reference curves and therefore cannot be rejected by a
Komolgorov-Smirnov test (95\% confidence); see the bottom left panel of Figure
\ref{fig:results:NSNS}.  Therefore 35 out of 500, roughly $7\%$ of 
all models, appear fully consistent 
 with all observations considered here and with the assumption that short
  GRBs are produced exclusively by all NS-NS mergers.

Because observations of the  Milky
Way suggest a NS-NS merger rate towards the high end of what our
simulations produce (see the bottom panel of Fig.
\ref{fig:results:MergerRates}), and because spiral star formation
extends through the recent universe -- that is, precisely into the
time intervals during which a significant fraction of short GRBs have
been observed (see Figure \ref{fig:data:grbs}) --  to an
excellent approximation we can conclude that, if short GRBs are due to NS-NS
mergers,  the best-fitting models of all
observations \emph{produce mergers and short GRBs preferentially in
  spiral galaxies}, with rapid mergers following quickly upon recent
star formation.   More specifically, we can generically draw the
following explicit conclusions:

\noindent \emph{High spiral mass efficiency needed}:  As indicated in
 Figure \ref{fig:results:MergerRates}, only a few populaton synthesis
 simulations can produce as many merging NS-NS binaries in spiral
 galaxies as does a natural extrapolation of the known Milky Way NS-NS merger rate.
The few ``model
universes'' which reproduce these high merger rates necessarily have
non-typical high mass efficiencies $\lambda_s$ for forming double
neutron stars in spiral galaxies.

\noindent \emph{High spiral fraction preferred}: As a consequence of
such high mass efficiencies $\lambda_s$ -- that is, because of the
high rate at which the spiral galaxies in these ``universes'' produce
NS-NS systems --  most mergers occur in spirals; equivalently, the spiral fraction $f_s$ of
constraint-satisfying models is often strongly biased towards spiral
galaxies ($f_s>1/2$).  The bottom panel of Figure
\ref{fig:results:SpiralFraction} explicitly shows that, though
unconstrained ``universes'' can produce mergers
preferentially in either elliptical or spiral galaxies (solid curve),
in order to reproduce observations of the Milky Way's merging binary
pulsars a ``universe'' almost always has its mergers predominantly in spirals
(dotted curve).

Similarly, since  a significant fraction of observed short GRBs 
are seen at low redshifts 
(as opposed to during the epoch of elliptical galaxy assembly),
spiral-dominated models usually fit observations better than
elliptical-dominated ones, as the latter could fit only given an
unusually long preferred delay between binary birth and merger.  For
this reason, demonstrated by the dashed curve in the bottom panel of Figure
\ref{fig:results:SpiralFraction},  the ``universes'' which best fit
the short GRB redshift distribution -- that is, which have short GRBs
occurring relatively recently, during the epoch of spiral star
formation -- largely produce their short bursts in spiral galaxies.

\begin{figure}
\includegraphics[width=\columnwidth]{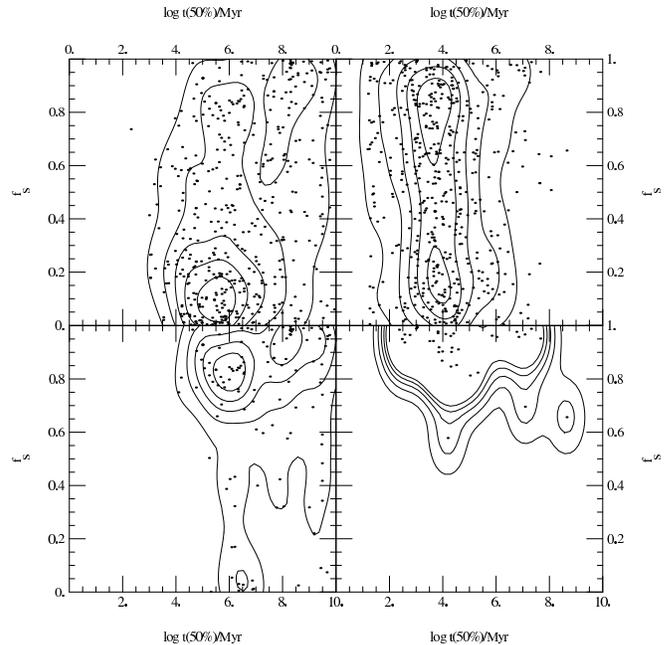}
\caption{\label{fig:results:EllipticalDelayContours} 
A mixed contour and scatter plot of the spiral fraction $f_s$ versus
the characteristic delay $t(50\%)$ between birth and merger in
\emph{elliptical} galaxies, for BH-NS
models  (left) and NS-NS models (right).  The top panels include all
500 models; the bottom panels provide only those models which satisfy
both constraints (for BH-NS simulations, the 137 ``model universes'' which are
 consistent with the short GRBs redshift distribution and detection
 rate; for NS-NS simulations, the 35 ``model universes'' which are
 consistent with the short GRB redshift distribution and the number of
 merging binary neutron stars in the Milky Way).
 This figure illustrates (i) that the vast majority of
 constraint-satisfying models have most mergers in spirals [also
 illustrated in Figure \ref{fig:results:SpiralFraction}] and more
 critically (ii) that the few models which permit most mergers to
 occur in elliptical galaxies are slightly biased towards
 longer characteristic delay times $t(50\%)$.
}
\end{figure}

\noindent \emph{Low spiral fraction implies long elliptical delays}:
Conversely, in order for \emph{elliptical} galaxies to host a
significant fraction of mergers, those elliptical galaxies must
produce binaries with an unusually long characteristic delay between
birth and merger.  For example, in the right two panels of  Figure
\ref{fig:results:EllipticalDelayContours} compare the prior (top) and
 constrained (bottom) distribution of spiral fraction and median delay
 time $t(50\%)$ between birth and merger.  As seen in the top
 panel, the most common delay between birth and merger in a randomly
 chosen elliptical population synthesis simulation is around $8
 \unit{Gyr}$; however, as seen in the bottom panel, for the handful of
 ``model universes'' which have fewer than $80\%$ of double neutron
 star mergers born in spiral galaxies, the median characteristic delay time
 is at least an order of magnitude larger.

\subsection{Comparisons II: BH-NS as burst source}
On the one hand, for purely technical reasons described in 
 \S~\ref{sec:results:full} we cannot require our BH-NS ``model
 universes'' to be consistent with the present-day NS-NS
 merger rate.  On the other hand,  unlike the NS-NS
 case, a few (84 out of 500) BH-NS ``model
 universes'' \emph{do} predict too few binary mergers  to reproduce
 observations, even given the most optimistic assumptions regarding
 beaming and the minimum luminosity of short GRBs; see 
Figure \ref{fig:results:DetectionRatesModFaintestBurst}.
Further,  as
shown in Figure \ref{fig:results:BHNS}, a significant fraction of redshift distributions
(183 out of 500, or $\simeq 37\%$) are consistent with the
limited sample, and many of those (137 out of 500, or 27\%) are
consistent with both the short GRB redshift distribution and detection
rate; see Figure \ref{fig:results:BHNS}.   From these constraints, we
can draw the following conclusions: 

\noindent \emph{Mergers usually in spirals}:
 Just as with NS-NS models, though \emph{a priori}
our ``model universes'' are equally likely to produce mergers and
short GRBs in elliptical and spiral galaxies, the set of ``model
universes'' which reproduce 
a short GRB redshift distribution that is dominated by recent mergers
is biased towards \emph{spiral-dominated models}: compare the solid
and dashed curves in the top panel of Figure \ref{fig:results:SpiralFraction}.
However, unlike the NS-NS case where both the redshift distribution
(preferring recent star formation)
and double neutron star merger rate (requiring very high merger rates
in spiral galaxies)  limit us to ``model universes''
that produce extremely many merging binaries, for BH-NS ``model
universes'' only the redshift distribution carries any information
that can bias results in favor of  spiral galaxies
[Fig. \ref{fig:results:SpiralFraction}].  In particular, as
illustrated by Figure \ref{fig:results:MergerRates}, the distribution
of spiral-galaxy BH-NS merger rates remains nearly the same no matter
what constraints have been applied.

\noindent \emph{Ellipticals can be hosts only with significant
  characteristic delays}:  Further, as one expects on physical
grounds, elliptical galaxies can contain a significant fraction of
short GRBs and mergers only when unusually long characteristic delays
between birth and merger are involved; see the left two panels of
Figure \ref{fig:results:EllipticalDelayContours}.
%

Unlike the double neutron star case, a significant
fraction of ``model universes'' predict \emph{low} spiral fractions
$f_s<0.8$.    In order for old elliptical 
galaxies to host a number of mergers similar to the young spiral
galaxies, the binaries in old elliptical galaxies must survive
over rather long times -- many Gyr after their formation.  
Comparatively speaking, population synthesis simulations of BH-NS
binaries produce more systems with the required characteristic long
delay time than do simulations of NS-NS binaries; see for example the
fraction of simulations in the top right panels in either 
Fig.  \ref{fig:popsyn:distrib:correlations} or the top left panel of 
\ref{fig:results:EllipticalDelayContours} with 
$t(50\%)$ greater than $10 \unit{Gyr}$.

\subsection{What fraction of mergers  produce short GRBs?}

So far, we have required our ``model universes'' to produce when
given the most favorable assumptions regarding burst luminosity
distribution and beaming  at least
as frequent short GRB detections as are observed, as shown in Figure 
\ref{fig:results:DetectionRatesModFaintestBurst}.  This fairly weak
constraint is imposed because, under
perfectly plausible but less optimistic assumptions, a great many short
GRBs could be missed.    
However, the difference in  Figure 
\ref{fig:results:DetectionRatesModFaintestBurst} between, on the one
hand, the vertical lines indicating the observed short GRB detection
rate and, on the other hand, an optimistic prediction $f_d f_b R_D$
can be reinterpreted as the \emph{fraction of short GRBs that
  must be missed} in order for our predictions to correspond with
observations. 

 By way of example, if a NS-NS ``model universe''
corresponds to  an optimistic detection rate of order
$10^4\unit{yr}^{-1}$, then only $1$ of every 100 NS-NS
mergers could produce short GRBs brighter than the least luminous seen
and aimed in our direction.    
This factor of $1/100$ could be produced by any one or combination of
several factors, all of which act to decrease the predicted short GRB
detection rate below these optimistic predictions:
(i) beaming, since the predicted detection rate is proportional to
$f_b$ and our optimistic calculations assume $f_b=1$;
(ii) fainter short GRBs, since the detection rate is also proportional
to  the
photon luminosity $\dot{N}_{\rm min}$ of the faintest short burst
(i.e., $\propto \dot{N}_{\rm min}/\dot{N}_{\rm min\, seen} $); 
(iii) some intrinsic physics which prevents all but a select few
mergers to produce detectable bursts; or even
(iv) further changes in our population synthesis model, such as 
assuming fewer than 100\% (our present choice) of all stars are born
in binaries.

No  incontrovertible evidence exists that
\emph{requires} any of these factors be less than unity.   However,
there is good observational and theoretical motivation for
re-evaluating our detection rate constraint using a prior on the
\emph{product} of all these factors: essentially, while all could be
nearly unity, good reasons exist to imagine that several may be
significantly less than 1.  
For example, (i) \citet{grb-051221a-host} and others have argued that
short GRBs may be beamed.  Further, (ii) the least luminous burst seen is
nearly at our detector's sensitivity limit, yet short GRBs peak flux
distribution is a featureless power law.  Since only a remarkable coincidence
could produce such a fortuitous combination of detector design and
source strength that this burst is indeed at or near the intrinsic
burst sensitivity limit, we can reasonably expect the minimum-luminosity burst to
be significantly less luminous than the faintest burst yet seen.
Additionally, as   described in
\citet{ChrisSpinup2007}, 
(iii)  theoretical simulations of BH-NS mergers  could possibly
produce short GRB merger events only for a limited array of binary component 
masses and spins; the fraction of mergers which could be short GRBs
ranges from 1/3 to 1/50, depending on the BH birth spin.
Finally, (iv) based on comparison with the present-day Milky Way, the
binary fraction could be up to a factor two lower than the value we
assume; e.g., \citet{ChrisSpinup2007} adopts a binary
fraction of 50\%.
If each of these four factors is  decreases the fraction of short GRB
detections below our predictions by only factor of roughly $3$, then
 our expectations for the short GRB detection rate corresponding to a
 given ``model universe'' should be lower by nearly a factor of
100!  

Adopting a canonical factor of 1/100 to transform the optimistic
detection rates presented in Figure
\ref{fig:results:DetectionRatesModFaintestBurst} into a ``realistic''
expectations leads to a dramatic transformation of our understanding.
On the one hand, extremely few
BH-NS ``model universes'' would produce short GRB events as frequently
as are observed.  On the other, these ``realistic'' assumptions
cause the 35
constraint-satisfying NS-NS ``model universes'' to predict roughly  as
many short GRB events as are observed (i.e., imagine shifting the
dotted peak in Figure
\ref{fig:results:DetectionRatesModFaintestBurst} to the left by 2
orders of magnitude).
While this tantalizing correspondance could be a coincidence and while
the precise results of our  ``realistic'' prior  cannot be taken too
seriously, they do remind us of two salient features of our
predictions: (i) that BH-NS ``model universes'' are consistent with
the data only given relatively optimistic assumptions, and that they
quickly become inconsistent with observations if those assumptions are
relaxed; and (ii) that on the other hand because  NS-NS ``model
universes'' can allow for so many detections and therefore have much
more flexibility  in the fraction of short GRBs that are missed, and
because double neutron star models are otherwise entirely consistent
with all existing observations, 
double neutron star models for short GRBs remain an attractive
candidate explanation for short GRBs.

\noindent \emph{Comparison to related studies}:
\citet{ChrisSpinup2007}  (hereinafter \abbrvChrisSpinup) also used the \texttt{StarTrack}
population synthesis code and hydrodynamical simulations of BH-NS
mergers with BH spins \citep{ManouBHNSMergers2007} to determine whether BH-NS merger events
occurred frequently enough, with the right combinations of
parameters,  to reproduce short GRB observations.   
In contrast to the present broad study, which relies only on
observational constraints, this study adopts several of the ``realistic''
priors mentioned above to reduce the fraction of merger events that
produce short GRBs.  Specifically, this paper relies on well-motivated
hydrodynamical studies to argue that at best a small fraction of BH-NS
mergers ($1/3$ to $1/50$, depending on black hole birth spin) will
produce  burst events.
Additionally, the \abbrvChrisSpinup{} simulations rely on a single
``most-plausible'' set of population synthesis parameters, including in particular a high
common envelope efficiency $\alpha\lambda$\footnote{The common envelope
efficiency reflects the fraction of orbital energy needed to eject the
envelope; since most BH-NS binaries go through a common-envelope
phase, a low efficiency implies these binaries will have a tight final
orbit.}.  As a result, compared to the wide range of common-envelope efficiencies
used in this study, \abbrvChrisSpinup\ find relatively low BH-NS merger
rates.
We also note that \abbrvChrisSpinup\ assume a 50\% binary fraction, lower
than the 100\% used here.
Combining these three factors,\footnote{Additionally, the simulations
  used in this paper allowed much less mass accretion onto black holes
  during the common-envelope phase.  Based on studies conducted in our
  group, this change does not produce a dramatic difference in merger
  \emph{rates}.}
\abbrvChrisSpinup\ conclude that BH-NS mergers
occur \emph{too infrequently} (less than
$10\unit{Gpc}^{-3}\unit{yr}^{-1}$) to explain short GRB merger rates.

\section{Conclusions}
\label{sec:conclude}

In this paper, we used a large archive of concrete, current
population synthesis calculations to generate merger rate densities
for NS-NS and BH-NS binaries.  Using assumptions regarding short
GRB source luminosities and detection selection effects,  we then compared these
rate densities to short GRB detection rates and redshift
distributions, as well as to (when available) the present-day Milky
Way binary neutron star merger rate.
Whether assuming short GRBs
arose from either BH-NS mergers or NS-NS mergers, 
a small but still substantial fraction of models were consistent with
existing observations
using the most optimistic assumptions for certain priors (i.e., no
beaming and all stars born in binaries).   
Contrary to an earlier study by \citet{Nakar}, we have demonstrated by
using a \emph{two-component} star formation model that
exceptionally long characteristic delays between binary birth and
merger are \emph{not} uniquely required needed 
to reproduce  the short GRB redshift distribution and detection rate,
though they are strongly preferred if we additionally demand a
significant fraction of short GRBs occur in elliptical hosts.
Generally our simulations favor \emph{spiral-dominated models}: $f_{s}>0.7$.   Double neutron star
models reproduce observations only if spiral hosts dominate; BH-NS
models can permit a wide range of spiral fractions.  
Though we have not imposed it as a constraint, the fraction of short
GRBs in spiral hosts potentially provides an extremely powerful
additional constraint on source models. 
For example,  the six host identifications shown in Table
\ref{tab:grbs} suggest the observed spiral fraction $\hat{f}_s$ is $50\%$.
 However, to impose it
reliably requires a careful study of the systematic uncertainties in
the two-component star formation model used  as well as of the
selection effects in host galaxy identifications.  
For example, more massive elliptical galaxies will more efficiently
retain any strongly-kicked neutron star binaries  but have  less gas into
which a the short GRB blast wave can collide and produce an afterglow;
these and other selection effects

Generally speaking, the exceedingly small number of well-identified short GRBs
strongly limits our ability to draw conclusions based on comparisons
to properties of that set,
such as to the redshift distribution or to the fraction of bursts seen
in spiral hosts.
More well-identified hosts are needed; additionally, these hosts
should hopefully be drawn from a less-biased sample than the present
sample appears to be  \citep[see][]{berger-manyfainthosts2006}.
Of course, we encourage deep follow-up searches for afterglows,
particularly since our model universes almost always predict a higher
proportion of  high-redshift short GRBs than has yet been observed.
We particularly encourage the development of thorough
\emph{redshift-limited} surveys, since a determination of the relative
proportion of bursts in star-forming or elliptical hosts can be done
in the nearby universe has significant potential to improve our
understanding of the formation mechanism. 

Though the limited sample of well-characterized short GRBs currently limits
our ability to constrain input physics, we expect that the significant
increase in event statistics expected over the next few years will
make these events a primary mechanism to constrain double compact
object merger rates and the associated astrophysics.
For example, with well-understood short GRBs already outnumbering the few known double
neutron stars in 
our galaxy, short GRBs would soon provide the most precise
(nonparametric) observational constraint on compact object merger
rates \citep[cf][]{Chunglee-nsns-1}, provided the selection effects
are quantitatively understood.  In turn, these constraints
inform us about  binary stellar evolution
(\cite{PSmoreconstraints},\cite{StarTrack2}).   
Additionally, since short GRB sources would also be copious emitters of
gravitational waves, Swift and ground-based gravitational-wave
observatories 
operating in coincidence could conceivably probe the
details of an unusually close merger event itself
\citep[see,e.g.,][]{2003ApJ...589..861K,2004ApJ...607..384F,Nakar}.  LIGO in
particular is both already operating at design sensitivity and
conducting triggered searches from GRB observations \citep{2005PhRvD..72d2002A}.
Given its compelling qualitative agreement and far-reaching scientific
impact, the merger hypothesis deserves a detailed comparison against the best
possible models, to either corroborate it, definitively disprove it,
or discover weaknesses in 
our assumptions and models.

Finally,  the current sample does not yet allow us to quantify the minimum luminosity and beaming of short GRBs. However assumptions  regarding the fraction of
merger events which do not produce short GRBs brighter than the
weakest burst seen so far and pointing towards us combined with 
theoretical priors \citep[see,e.g.][]{ChrisSpinup2007}, indicate that BH-NS mergers do not
occur frequently enough to explain all short GRBs. We anticipate that future larger samples will allow us to place stronger constraints on the merger models.

\begin{acknowledgments}
{\it Acknowledgments---}  We would like to thank Don Lamb, Ehud
Nakar, Avishay Gal-Yam, Re'eem Sari, Arieh Konigl,  Shri Kulkarni, Neil
Gehrels, and the participants of the Ringberg Short GRB conference of
2007 for helpful comments during
this paper's long gestation. This work was partially supported by NSF
award PHY-0353111 and a Packard Fellowship in Science and Engineering
awarded to VK, and grants KBN  1P03D02228 and 1P03D00530 to KB. 
\end{acknowledgments}

\bibliography{%
grb-detectors,long-grbs,short-grb,short-grb-sgr,short-grb-data,short-grb-data-analysis,short-grb-mergermodel,%
short-grb-data-analysis-jetbreaks,%
popsyn,popsyn-sed,popsyn_gw-merger-rates,%
star-formation-history,star-formation-properties,%
astrophysics-stellar-dynamics-theory,%
observations-galaxies-distributions,observations-clusters,galaxy-formation-theory,cosmology,LIGO-publications,gw-astronomy-bursts,observations-supernovae,observations-pulsars-kicks,observations-binaries-constraintsOnInteractions,%
mm-general,mm-statistics}

\appendix

\section{Archive selection}
\label{ap:archiveSelection}
\begin{figure}
\begin{centering}
\includegraphics[width=0.8\textwidth]{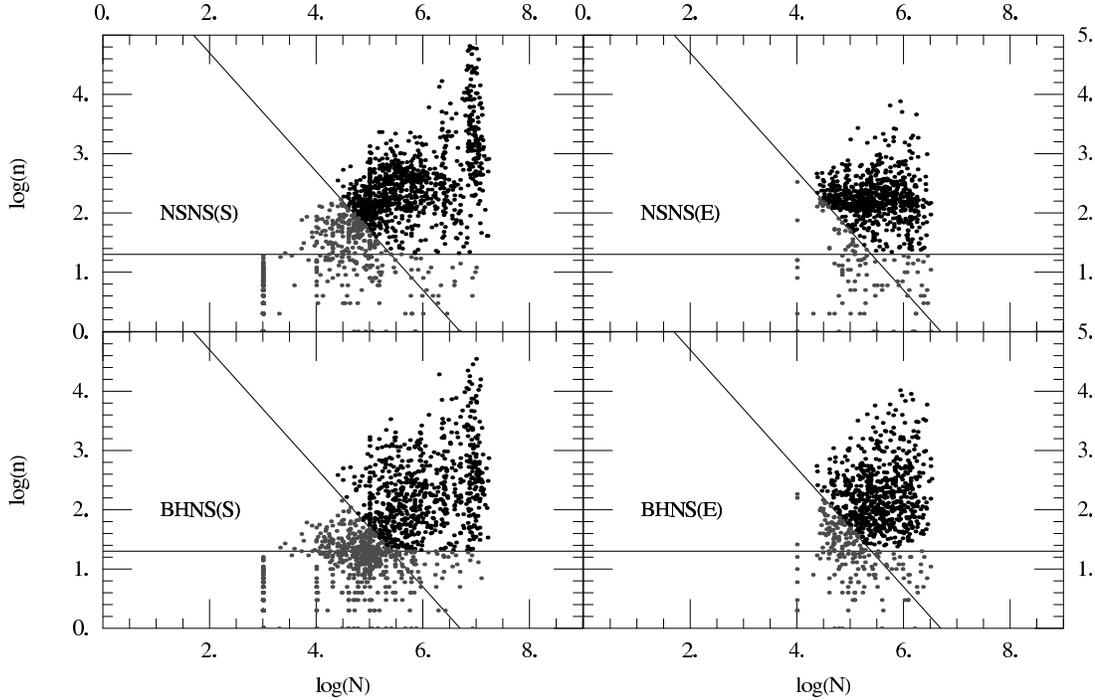}
\end{centering}
\caption{\label{fig:archiveselect:scatter}  For spiral (left two panels) and
  elliptical (right to panels) archives, a scatter plot of the number of NS-NS
  (top) and BH-NS (bottom)
  binaries $n$ seen in a simulation of $N$ progenitor binaries (all of
  $m>4 M_\odot$, with primary mass drawn from the broken-Kroupa IMF).
  The two solid lines show the cutoffs $n>\mynmin$ and $n N>\myratio$ imposed to
  insure data quality and reduce sampling bias.  Simulations used in
  this paper, shown in black, must lie above and to the right of these cutoffs.
}
\end{figure}

The predicted short GRB detection rate depends directly on the
fraction of star forming mass $\lambda$ predicted to end up in GRB
progenitors.  However, the mass efficiency $\lambda$  can vary
substantially depending on  model assumptions.  In order to constrain
the \emph{range} of short GRB detection rates expected (for a fixed
minimum short GRB luminosity $\dot{N}$), we require the unbiased
\emph{distributrion} of $\lambda$.
However, the large variety of simulation sizes in our simulations
introduces a bias in our estimate of the mass efficiency distribution
-- or, equivalently, in the distribution of $n/N$ for $n$ the number
of binaries seen and $N$ the number of binaries simulated.  
Specifically, not all of our archived population synthesis simulations contain
enough of each type of event (indexed by $K$) to produce reliable
predictions involving it.  The set of simulations with $n$ greater
than  \emph{any fixed threshold} threshold
(including $n=0$) is biased,  over-representing simulations with high $n/N$; see for example Figure
\ref{fig:archiveselect:scatter}.   
Since the mass efficiency $\lambda$ is directly proportional to $n/N$
[Eq.~(\ref{eq:popsyn:lambda})], we describe a filter on $n$ and $N$
which preserves the distribution of $n/N$  
(for the distribution of $\lambda$)
and \emph{simultaneously}
insures that the average simulation has many binaries
($\left<n\right> > 200$, so each model's delay time distribution $dP/dt$ can be accurately estimated).

Formally, each population synthesis simulation converts some unknown
fraction $s$ of progenitor binaries into target events $n=s N$.  In
practice our population sample size $N$ is roughly randomly chosen,
independent of the unknown $s$.  In other words, we expect the
distribution of $n$ and $N$ to derive from two independent
distributions for $N$ and $s$, with densities $f$ and $g$:
\begin{eqnarray}
dP (\ln N, \ln n) &=& f(\ln N) g(\ln (n/N)) d\ln N d\ln n \nonumber \\ 
 &=& f(\ln N) g(\ln n/N) d \ln \sqrt{N n}  d \ln n/N
\end{eqnarray}
The distribution of $s=n/N$ can be quickly extracted from any 
 two-dimensional distribution in $n,N$ by:
\begin{eqnarray}
g d \ln s = \int \delta\left(\ln(s/(n/N))\right)  d P\; .
\end{eqnarray}
However, the above method assumes the distribution perfectly
resolved.  In practice simulations are generally \emph{not} repeated,
so ``fractional'' and small  $n$ cannot be resolved by repeated trials; instead,
only simulations with $s\gg 1/N$ will produce enough events to provide
reliable estimates of $s$, and thus its distribution.  This truncation
effect biases our reconstruction of the two-dimensional density and
therefore our reconstruction of the mass efficiency distribution.
If we had perfect foreknowledge, however, we could have chosen our
sample size so $n N$ was constant and large.  Though we would choose $n$ and $N$  in
a highly correlated fashion, we would guarantee each point was
well-resolved; our method above would correctly reconstruct the
distribution of $s$.   
Hence choosing all population synthesis archives with $n N$ greater
than any threshold will allow accurate estimation of the distribution
of $s$.

Based on the above discussion and the observed distributions in $n$
and $N$, we introduce cutoffs on $n$ ($>\mynmin$) and $n N$ 
($>\myratio$)
which remove the least resolved simulations from consideration 
(via the $n$ cutoff), reduce the bias associated with a minimum $n$
(via the $n N$ cutoff), and insure that most simulations have at least
$200$ binaries (both).


\section{Estimating Delay Time Distribution}
\label{ap:smoothingSystematics}
Poisson errors associated with the limited sample  size of our
population synthesis simulations 
inhibits our ability to reconstruct the delay time distribution,
whether represented as a cumulative distribution $P_m(t)\equiv P_m(<t)$,
the probability that a delay between birth and merger is less than
$t$, or as $dP_m/dt$.  
Using classical statistical methods -- see, e.g., \citet{Merritt1994},
\citet{ThomsonTapia}, and references therein --  we smooth the $n$ observed
delay times to build our estimates for $P_m(t)$ and $dP_m/dt$.  This
appendix briefly reviews those methods and the accuracy of the
resulting estimates.

\noindent \emph{Estimating cumulative distribution}: We estimate the cumulative
distribution $P(<t)$ for a sample of population synthesis events by a
function $\hat{P}(t)$ that smoothes sample dataover a short 
smoothing length $s$ in $l_t\equiv \log_{10} (t/{\rm Myr})$:
\begin{eqnarray}
\hat{P}_s(l_t)&\equiv& \sum_{k=1}^{n} \Theta_s(l_t - l_{t,k}) \\
\Theta_s(x) &=& \frac{1}{2}\left(1+{\rm erf}(x/s\sqrt{2})\right) \\
s_1 &=& \frac{\left[({\rm max}_k l_{t,k}) - ({\rm min}_k
    l_{t,k})\right]}{10 n^{0.4}}
\end{eqnarray}
for $n$ the number of events in our population synthesis sample,
$k=1\ldots n$ an index over tat same sample,  and
$l_{t,k}=\log (t_k/{\rm Myr})$ the logarithm of the delay $t_k$
between binary formation and merger for the $k$th binary.
The extremely short smoothing length used here approximately
minimizes the average difference between predictions and results.

\begin{figure}
\includegraphics{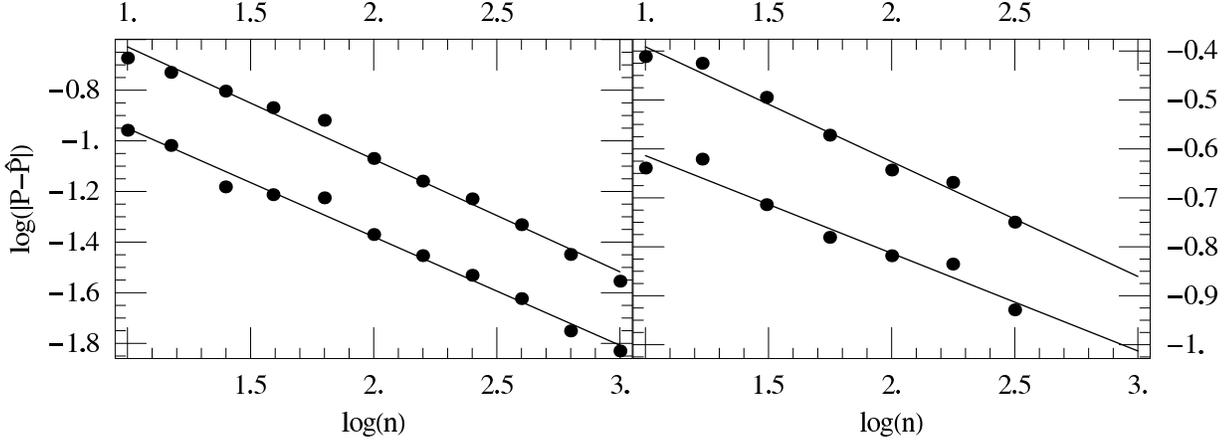}
\caption{\label{fig:fits:PcumSmoothingWorks} Results for errors in
  estimates of  $P(t_m)$ and $dP_m/dt$, versus the number of points used
  in the estimate, for the trial distribution function $dP_m/dt=
  \frac{0.079}{t}$ between $30 $ and $10^7\unit{Myr}$. 
 Left: Plot of the average and
  $90\%$ supremum-norm
  error  versus $n$ for $P_m(t)$, along with fits to these two
  quantities (namely, $n^{-0.45}/3$ and $0.7n^{-0.47}$,
  respectively).
Right: Plot of the average and 90\% bound on fractional error in
$dP_m/dt$ versus the number of  points available, along with fits to
these two quantities (namely, $0.36 n^{-0.17}$ and $0.73n^{-0.23}$, respectively).
}
\end{figure}

To test this approach, for several $n$ we drew many
Monte Carlo samples of $n$ delay times from a   canonical cumulative
distribution:
\[
P_m(l_t) = \frac{lt-\log(30)}{7-\log(30)}
\Theta(l_t-\log(30))\Theta(7-l_t) 
\]
Figure \ref{fig:fits:PcumSmoothingWorks} shows the average and 90\% distance ${\rm max}_{l_t}
|P_m(l_t)-\hat{P}_m(l_t)|$ between our function and our fit versus the
number of points smoothed to estimate $P_m(t)$.  
%
%

Given the archive selection procedure presented in
\ref{ap:archiveSelection}, each archive typically contains of order
$n\approx 100$ merging binaries.  For such a typical archive, our
smoothing method will reconstruct 
$P(t)$ almost everywhere to better than $5\%$, with the largest errors
typically arising near the largest and smallest delay times.
As expected, the maximum error agrees perfectly with the distance
between observations and data for a Kolmogorov-Smirnov 95\% hypothesis test.

\noindent \emph{Estimating differential distribution} :  Similarly, we
estimate $dP_m/dt$ by $\hat{\dot{P}}_{m,10 s}$, defined using the time derivative
of $P_m(<t)$  after
converting from logarithmic to physical time:
\begin{eqnarray}
\hat{\dot{P}}_m &\equiv& \frac{1}{t \ln 10}\frac{d\hat{P}_m}{dl_t} =
    \frac{1}{t \ln 10}\sum_{k=1}^{n} \frac{1}{\sqrt{2\pi(s_2)^2}}
       e^{-(\log t - l_{t,k})^2/2(s_2)^2} \\
s_2 &\equiv& \frac{\left[({\rm max}_k l_{t,k}) - ({\rm min}_k
    l_{t,k})\right]}{1.25\sqrt{n}}
\end{eqnarray}
Compared to the previous case of estimating the \emph{cumulative}
distribution $P_m(t)$, where a significant fraction of all sampled
points affect the estimate at any $t$, a significantly longer
smoothing length $s_2$ is required to estimate $dP_m/dt$, since roughly
only those few sample points within $s_2$ of $t$ contribute to our estimate.
As a result, we cannot confidently estimate $dP_m/dt$ more accurately than $\approx
30\%$.

To demonstrate that smoothing produces a fairly inaccurate prediction
for $dP_m/dt$, we applied it to the trial
problem mentioned above.   To 
estimate the mean relative error $I$ associated with our estimate in the
\emph{physically pertinent interval} of roughly $100$ Myr to $15$ Gyr by
\begin{eqnarray}
I&\equiv& \left[\int_{1.8}^{4.5} \frac{dl_t}{2.7} 
  \frac{|\hat{\dot{P}}-\dot{P}|^2}{(\hat{\dot{P}}^2+\dot{P}^2)/2} 
 \right]^{1/2} \; ,
\end{eqnarray}
which generalizes an rms measurement of the relative error in $P$ to
allow for large relative errors
 [i.e., when  $\hat{\dot{P}}_m=\dot{P}_m(1+\delta(l_t))$,
then $I =  (\int \delta^2 dl_t/\Delta l_t)^{1/2}$].  The second panel of Figure
\ref{fig:fits:PcumSmoothingWorks} demonstrates that large errors are
inevitable, even with many sample points and using a smoothing length
$s_2$ which is approximately optimized for each $n$. 
While this slow convergence is
a familiar problem for all density estimators
\citep[see,e.g.][]{ThomsonTapia}, it severely  limits the ability of
any population synthesis simulation to reliable estimate
$dP_m/dt$, given the severe computational limits involved.  By way of
example, a delay time distribution accurate to $5\%$ would require
roughly $6^4\approx 1300$ times more binaries than a $30\%$ accurate estimate.

\noindent \emph{Implications for short GRB predictions}: The short GRB
detection rate, redshift distribution, and elliptical-to-spiral ratio
all depend on estimates of $dP_m/dt$.  
However, while our reconstruction of $dP_m/dt$ for any \emph{fixed}
population synthesis model involves substantial
uncertainties, these are not much greater than our corresponding
uncertainty in our estimate of that model's the mass efficiency
$\lambda$.  Further, these uncertainties are vastly \emph{less} than
the systematic uncertainty involved in not knowing the physically
appropriate population synthesis model (i.e., roughly two orders of
magnitude uncertainty in $\lambda$).

\end{document}

%% file: publishable-table.tex
\begin{deluxetable*}{llllllllll}
\centering\tablecolumns{10}\tablecaption{Short Gamma Ray Bursts}\tabletypesize{\footnotesize}\tablehead{
\colhead{ GRB$^a$}&
\colhead{ Det$^b$}&
\colhead{ z$^c$}&
\colhead{ T90$^d$}&
\colhead{ P$^e$}&
\colhead{ Id$^f$}&
\colhead{ OA$^g$}&
\colhead{ Type$^h$}&
\colhead{ Usage$^i$}&
\colhead{ Refs $^j$}
}
 \startdata
 050202  & S  & -  & 0.8  & -  & F  & -  & -  & S3  &1 \\ 
 050509B  & S  & 0.226  & 0.04  & 1.57  & T  & F  & E  & S1  &2,3,4,5,6 \\ 
 050709  & H  & 0.161  & 0.07  & 0.832  & T  & T  & S  & SH1  &7,8,9,10,11,6,12 \\ 
 050724  & SH  & 0.257  & 3.  & 3.9  & T  & T  & E  & S1  &7,13,14,15,16,1,6 \\ 
 050813  & S  & 1.8  & 0.6  & 1.22  & T  & F  & -  & S1  &17,5,1,6 \\ 
 050906  & S  & -  & 0.128  & 0.22  & F  & F  & -  & S3  &1 \\ 
 050925  & S  & -  & 0.068  & -  & F  & -  & -  & S3  &6 \\ 
 051105A  & S  & -  & 0.028  & -  & F  & -  & -  & S3  &1 \\ 
 051114A  & S  & -  & 2.2  & -  & F  & -  & -  & S3  &18 \\ 
 051210  & S  & z $>$ 1.55  & 1.2  & 0.75  & T  & F  & -  & S2  &19,1,20,21,6 \\ 
 051211A  & H  & -  & 4.25  & -  & F  & -  & -  & SH3  &1 \\ 
 051221A  & S  & 0.547  & 1.4  & 12.1  & T  & T  & S  & S1  &22,1,21,6 \\ 
 060121  & H  & 4.6,1.5  & 1.97  & 4.93  & T  & T  & -  & SH2  &23,1,21,24 \\ 
 060313  & S  & z $<$ 1.7  & 0.7  & 12.1  & T  & T  & -  & S2  &25,1,21,6 \\ 
 060502B  & S  & 0.287  & 0.09  & 4.4  & F  & F  & E  & S1  &26,1,21,6 \\ 
 060505  & S  & 0.089  & 4.  & 1.9  & T  & T  & S  & S1  &1,27,28 \\ 
 060801  & S  & 1.13  & 0.5  & 1.3  & T  & F  & -  & S1  &21 \\ 
 061006  & S  & 0.438  & 0.42  & 5.36  & T  & T  & -  & S1  &21 \\ 
 061201  & S  & 0.1,0.237  & 0.8  & 3.9  & T  & T  & -  & S2  &6 \\ 
 061210  & S  & 0.41  & 0.19  & 5.3  & T  & T  & -  & S1  &6 \\ 
 061217  & S  & 0.827  & 0.212  & 1.3  & T  & T  & -  & S1  &6 \\ 
\enddata\label{tab:grbs}
\tablenotetext{1}{Gamma-ray burst index
}
\tablenotetext{2}{Detector in which the GRB was initially detected; S denotes Swift, H denotes HETE-II.
}
\tablenotetext{3}{Redshift of the host, if well identified.
}
\tablenotetext{4}{Duration of the burst.
}
\tablenotetext{5}{Peak photon flux of the burst (ph/cm$^2$/s).
}
\tablenotetext{6}{Whether the host was optically identified.
}
\tablenotetext{7}{Whether the burst produced a visible optical afterglow.
}
\tablenotetext{8}{Morphology of the host: elliptical (E) or spiral (S).
}
\tablenotetext{9}{Summary of the previous columns: S1 bursts were initially seen by Swift and have a well-defined redshift; S2 bursts were seen by Swift and have some uncertain redshift information; S3 bursts include all bursts seen by Swift only.  Similarly, SH1 includes all bursts seen by Swift or HETE-2 with a well-defined redshift. 
}
\tablenotetext{10}{References:
(1) \cite{DonLamb-modifiedClassifications2006}
(2) \cite{grb-050509b-discovery-groupA}
(3) \cite{2005ApJ...630L.165L}
(4) \cite{grb-050509b-host-2006}
(5) \cite{2005ApJ...634..501B}
(6) \cite{berger-shortgrb-parameter-correlation-2007}
(7) \cite{grb-050709-discovery}
(8) \cite{2005Natur.437..845F}
(9) \cite{grb-050709-discovery-groupB}
(10) \cite{2005ApJ...634L..17P}
(11) \cite{2006AA...447L...5C}
(12) \cite{Gehrels-shortgrb-SwiftReview-Mid2007}
(13) \cite{grb-050724-host}
(14) \cite{2006ApJ...642..989P}
(15) \cite{2006AA...454..113C}
(16) \cite{2006astro.ph..3773G}
(17) \cite{Berger2006talk}
(18) \cite{NakarReviewArticle2006}
(19) \cite{2006AA...454..753L}
(20) \cite{grb-compilation-LIGO-inspiral}
(21) \cite{berger-manyfainthosts2006}
(22) \cite{grb-051221a-host}
(23) \cite{grb-060121-discovery-afterglow}
(24) \cite{2006ApJ...648L..83D}
(25) \cite{2006astro.ph..05005}
(26) \cite{2006ApJ...638..354B}
(27) \cite{grb-060505-argue1}
(28) \cite{grb-060505-argue2}
}
\end{deluxetable*}